\documentclass[%
 reprint,
superscriptaddress,
 amsmath,amssymb,
 aps,
prb,
]{revtex4-2}
\usepackage{color}
\usepackage[pdftex,colorlinks=true,linkcolor=red,citecolor=blue]{hyperref}
\usepackage{graphicx}
\usepackage{dcolumn}
\usepackage{bm}
\usepackage{ulem}

\begin{document}


\title{Competition of Magnetocrystalline Anisotropy of Uranium Layers and Zig-Zag Chains in UNi$_{0.34}$Ge$_2$ Single Crystals}

\author{Adam P. Pikul}
\email{a.pikul@intibs.pl}
\address{Institute of Low Temperature and Structure Research, Polish Academy of Sciences, 50-422 Wroc{\l}aw, Poland}
\address{Idaho National Laboratory, Idaho Falls, Idaho 83415, USA}%
\author{Maria Szlawska}
\address{Institute of Low Temperature and Structure Research, Polish Academy of Sciences, 50-422 Wroc{\l}aw, Poland}
\author{Xiaxin Ding}
\address{Glenn T. Seaborg Institute, Idaho National Laboratory, Idaho Falls, ID 83415, USA}
\author{J{\'o}zef Sznajd}
\address{Institute of Low Temperature and Structure Research, Polish Academy of Sciences, 50-422 Wroc{\l}aw, Poland}
\author{Masashi Ohashi}
\address{Institute of Science and Engineering, Kanazawa University, Kakuma-machi, Kanazawa 920-1192, Japan}
\author{Dorota A. Kowalska}
\address{Institute of Low Temperature and Structure Research, Polish Academy of Sciences, 50-422 Wroc{\l}aw, Poland}
\author{Mathieu Pasturel}
\address{Univ Rennes, CNRS, Institut des Sciences Chimiques de Rennes, UMR6226, Rennes, France}
\author{Krzysztof Gofryk}
\address{Idaho National Laboratory, Idaho Falls, Idaho 83415, USA}%

\date{\today}

\begin{abstract}
Structural and thermodynamic properties of single-crystalline UNi$_{1-x}$Ge$_2$ with $x$\,=\,0.66 have been investigated by measuring magnetization, specific heat, and thermal expansion over a wide range of temperatures and magnetic fields. The measurements revealed the emergence of a long-range antiferromagnetic ordering of uranium magnetic moments below the N{\'e}el temperature $T_{\rm N}$\,=\,45.5(1)\,K and the existence of two easy axes in the studied compound, namely $b$ and $c$, which correspond to the plane of the uranium zig-zag chains. Magnetic field applied along these two crystallographic directions induces in the system a first-order metamagnetic phase transition (from antiferromagnetism to field-polarized paramagnetism), and the width of the magnetic hysteresis associated with that transition reaches as much as about 40 kOe at the lowest temperatures. A magnetic phase diagram developed from the experimental data showed that the metastable region associated with that magnetic hysteresis forms a funnel that narrows toward the N{\'e}el point in zero magnetic field. The four-layer Ising model has successfully predicted the collinear antiferromagnetic structure in UNi$_{0.34}$Ge$_2$ (known from earlier reports), its magnetic phase diagram, and temperature and field variations of its magnetization. Moreover, it suggests that the first-order phase transition extends down to zero magnetic field, although it is barely detectable in the experiments performed in low magnetic fields. According to this model, the second-order phase transition occurs in the compound only in zero field.
\end{abstract}


\maketitle



\section{Introduction}

Uranium intermetallics have attracted the attention of the scientific community not only because of their potential use in the nuclear industry as new accident tolerant fuel materials \cite{Kim2012,Ortega2016}, but also because of the unique and intriguing physical properties that have been discovered in many of them as a direct consequence of the presence of 5f shells in their electron structure. Those phenomena include but are not limited to heavy-fermion superconductivity \cite{Pfleiderer2009,Ran2019,Aoki2019a}, very high magnetic ordering temperatures and strong magnetocrystalline anisotropy \cite{Kaczorowski1990,Stalinski1974}, or recently discussed dual nature of the uranium 5f electrons \cite{Zwicknagl2003,Troc2012,Lee2018,Amorese2020}.

A unique place among uranium compounds is occupied by uranium germanides, as undoubtedly one of the most spectacular discoveries made among uranium compounds was the finding of pressure-induced superconductivity in ferromagnetically ordered UGe$_2$ (with the superconducting temperature reaching $T_{\rm sc}$\,=\,0.8\,K at pressure $P_{\rm sc}$\,=\,1.2\,GPa) \cite{Saxena2000}, followed shortly thereafter by the discovery of ambient-pressure superconductivity in the ferromagnets URhGe ($T_{\rm sc}$\,=\,0.25\,K) \cite{Aoki2001} and UCoGe ($T_{\rm sc}$\,=\,0.8\,K) \cite{Huy2007}. Importantly, the ferromagnetism of UGe$_2$ is relatively strong: the compound has a high Curie temperature $T_{\rm C}$\,=\,52\,K and very large (as for uranium intermetallic compounds) ordered magnetic moment $\mu_{\rm ord}$\,=\,1.5\,$\mu_{\rm B}$. In contrast, URhGe and UCoGe exhibit much weaker ferromagnetism: their $T_{\rm C}$ and $\mu_{\rm ord}$ are as small as -- respectively -- 9.5\,K and 0.42\,$\mu_{\rm B}$ in URhGe \cite{Aoki2001} and 3\,K and 0.03\,$\mu_{\rm B}$ in UCoGe \cite{Huy2007}, which is certainly not indifferent for the formation of a superconducting state in those systems.

The compounds U$T\!E_{1-x}$Ge$_2$ (where $T\!E$ stands for a transition metal) are another, relatively recently discovered and described in the literature, family of uranium germanides. They crystallize in the orthorhombic CeNiSi$_2$ structure type (in the case of the phases with $T\!E$ = Fe \cite{Henriques2015,Szlawska2022,Pikul2022}, Co \cite{Soude2010}, and Ni \cite{Molcanova2017,Ohashi2018,Pasturel2021}), or a slightly monoclinically distorted derivative (for $T\!E$ = Ru \cite{Pasturel2018} and Os \cite{Pikul2019JdA}). A common feature of the compounds U$T\!E_{1-x}$Ge$_2$ and the ferromagnetic superconductors UGe$_2$, URhGe, and UCoGe is the presence of parallel uranium zig-zag chains. Importantly, in the latter three phases, these chains induce in the ferromagnetically ordered state local inversion symmetry breaking, which seems to be of great importance for the formation of the superconducting state \cite{Aoki2019}. Therefore, although superconductivity has not yet been observed in any of the U$T\!E_{1-x}$Ge$_2$ phases (at least at ambient pressure and at temperatures above 2~K), their structural similarity to ferromagnetic superconductors warrants more in-depth studies of their properties.

So far, it has been found that most of the U$T\!E_{1-x}$Ge$_2$ compounds are ferromagnets with the Curie temperature ranging from 18~K (for Co \cite{Soude2010}) to 63~K (for Ru \cite{Pasturel2018}). Only one representative of that group of compounds, namely UNi$_{1-x}$Ge$_2$ has been found to order antiferromagnetically, with the N{\'e}el temperature $T_{\rm N}$\,=\,47~K (for $x$\,=\,0.55) \cite{Pasturel2021}. Furthermore, at the temperature of 2~K and in a magnetic field
of about 100~kOe, a first-order metamagnetic transition to field-polarized paramagnetism (often called field-induced ferromagnetism) has been observed in that compound, accompanied by a very large magnetic hysteresis with a width reaching about 40~kOe \cite{Pasturel2021}.

\begin{figure}[t]
\includegraphics[width=\columnwidth]{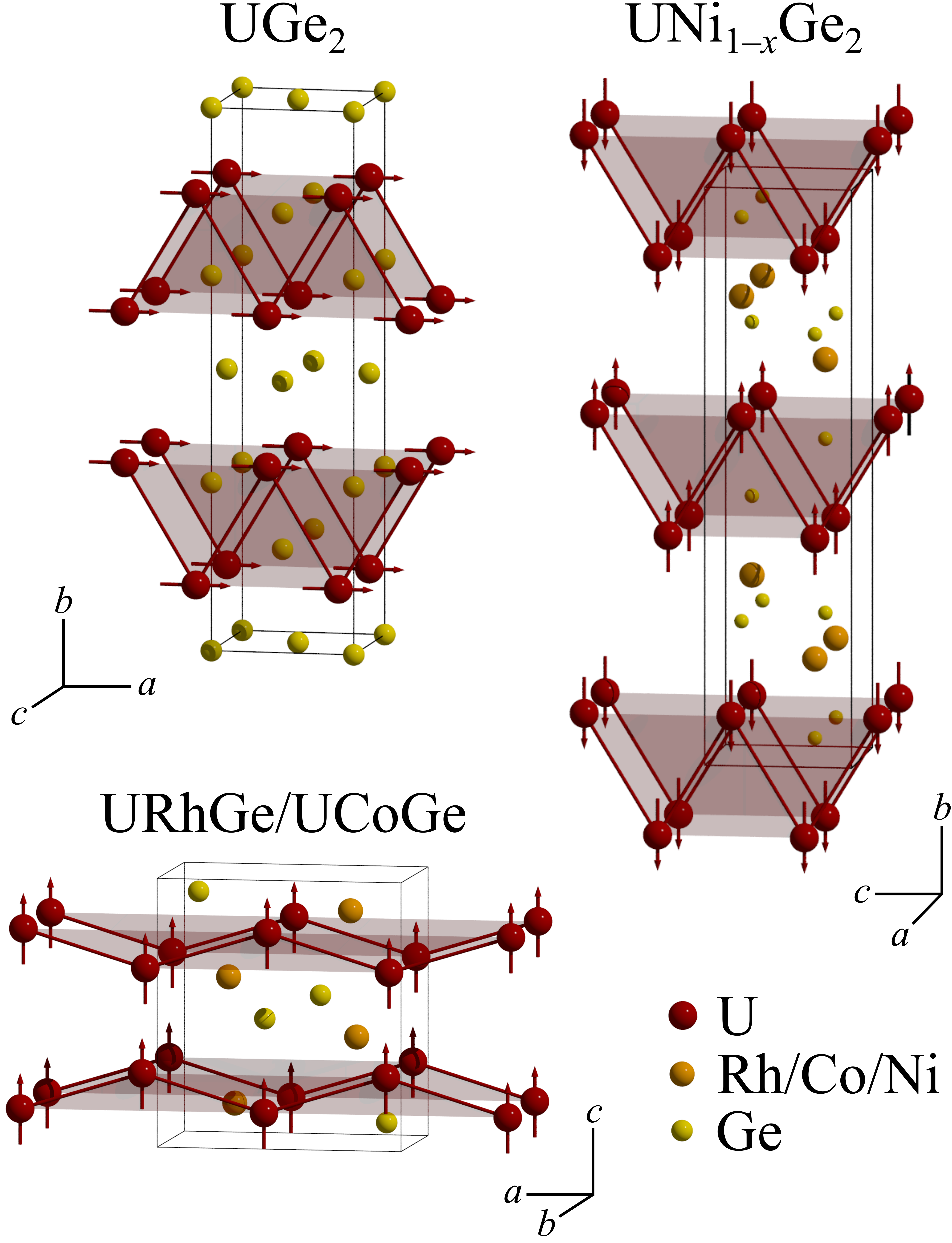}
\caption{\label{fig:cryst-struct}(Color online) Magnetic structures of UGe$_2$, U(Rh,Co)Ge and UNi$_{1-x}$Ge$_2$ in the ferromagnetic state and in the absence of external magnetic field, drawn from data taken from Refs.~\onlinecite{Aoki2019,Aoki2019,Pasturel2021}.}
\end{figure}

Neutron diffraction experiments performed on a polycrystalline sample of UNi$_{0.5}$Ge$_2$ \cite{Pasturel2021} revealed that in the absence of external magnetic field, the magnetic moments of uranium in that system lie in the plane of the chains and are aligned perpendicular to the zig-zag chains direction (see Fig.~\ref{fig:cryst-struct}). Moreover, they are ferromagnetically ordered within the chains (\textit{i.e.} along the crystallographic $b$-axis), while the consecutive pairs of chains are antiferromagnetically ordered and form the sequence $-++-$ \cite{Pasturel2021}. However, there is no information on the evolution of the magnetic structure with magnetic field, leading to the metamagnetic transition in that system.

As can be seen in Fig.~\ref{fig:cryst-struct}, the arrangement of the magnetic moments within the chains in UNi$_{1-x}$Ge$_2$ is very similar to that reported for URhGe and UCoGe \cite{Aoki2019}. This is somewhat surprising because the distances and angles between the uranium atoms within the chains in UNi$_{1-x}$Ge$_2$ are much closer to these found in UGe$_2$, where the ferromagnetically ordered magnetic moments are aligned not perpendicular to the chains but along them \cite{Aoki2019}. It is therefore of particular interest to study in detail the magnetic behavior of the system UNi$_{1-x}$Ge$_2$, including attempts to model the magnetic interactions between the uranium magnetic moments.

In this paper we focus on results of crystal structure (Sec.\,\ref{sec:cryst-struct}) and basic thermodynamic characterization (Sec.\,\ref{sec:magn-grnd-st}) of a single crystal of UNi$_{1-x}$Ge$_2$ with $x=0.66$. They are followed by a description of the results of experiments performed in magnetic fields devoted to detailed characterization of the metamagnetic phase transition evidenced in the studied compound (Sec.\,\ref{sec:metamagn-trans}). Finally, we construct a magnetic phase diagram and propose its theoretical description (Sec.\,\ref{sec:phase-diag}). We show that the magnetic behavior of the studied uranium germanide can be explained within a straightforward model that is independent of the crystal and band structure of this compound. We believe that our findings can help in the proper interpretation and understanding of the unique physical properties of the closely related ferromagnetic superconductors UGe$_2$, URhGe, and UCoGe.


\section{Methods}

A high-quality single crystal of UNi$_{1-x}$Ge$_2$ was grown using a tetra-arc furnace by the Czochralski-method pulling method. The obtained cylindrical-shaped crystal with a diameter of about 3\,mm and a length of about 15\,mm was then cut along the main axis of the cylinder and checked by X-ray diffraction (XRD), energy-dispersive X-ray spectroscopy (EDXS) and selected area electron diffraction (SAED). The single-crystal XRD experiment was performed on an Oxford Diffraction X'Calibur four-circle diffractometer equipped with a CCD camera using graphite-monochromated MoK$\alpha$ radiation ($\lambda$\,=\,0.71073\,\AA). Data collection was carried out at room temperature on a crystal with dimensions of about $0.16 \times 0.06 \times 0.03$\,mm$^3$ using the CrysAlis PRO 1.171.39.46 software (Rigaku Oxford Diffraction, 2018), which was also involved in data reduction. The empirical absorption was corrected by Gaussian integration over a multifaceted crystal model. The crystal structure was solved by direct methods and refined by the full-matrix least-squares method on F$^2$ utilizing the crystallographic software package SHELX \cite{Sheldrick2015} accessible through the graphical user interface Olex2 \cite{Dolomanov2009}. EDXS measurements were performed using a FESEM FEI Nova NanoSEM 230 scanning electron microscope equipped with an EDAX Genesis XM4 spectrometer on a cleaved surface of the single crystal. SAED experiments were carried out using a JEOL 2100 LaB$_6$ transmission electron microscope operating at 200~kV.

All further physical experiments were performed on specimens cut from the original single-crystalline cylinder. Crystallinity and orientation of those samples were checked by the Laue backscattering technique using a Proto Laue-COS Single-Crystal Orientation System. Longitudinal DC magnetization and specific heat of the so-obtained bar-shaped single crystal were measured down to about 2\,K and in applied magnetic fields up to 140\,kOe using a commercial Quantum Design Physical Property Measurement System (QD PPMS) equipped with a vibrating sample magnetometer (VSM). Thermal expansion and magnetostriction were studied over the same temperature and magnetic field ranges using a QD PPMS DynaCool platform equipped with a stress-dilatometer from Kuechler Innovative Measurement Technology.


\section{Results and discussion}
\subsection{\label{sec:cryst-struct}Crystal structure}

\begin{table}
\caption{ \label{tabSI1} Single crystal X-ray diffraction data collection and refinement parameters for UNi$_{0.34}$Ge$_2$.}
\centering
\begin{ruledtabular}
\begin{tabular}{lc}
 Empirical formula & UNi$_{0.34(1)}$Ge$_2$ \\ 
\hline
Structure-type & CeNiSi$_2$ \\
Molar mass & 403.02\,g\,mol$^{-1}$\\ 
Space group & $Cmcm$ \\
Cell parameters & $a$\,=\,4.1000(2)\,\AA\\ 
 & $b$\,=\,15.8711(9)\,\AA\\
 & $c$\,=\,4.0195(2)\,\AA\\
Cell volume & 261.55(2)\,\AA$^3$ \\
$Z$ / calculated density & 4 / 10.235\,g\,cm$^{-3}$ \\
Absorption coefficient & 86.563\,mm$^{-1}$ \\ 
Crystal colour & black  \\
Crystal size & 0.16\,$\times$\,0.056\,$\times$\,0.026\,mm$^3$ \\
$\theta$ range & 2.567$^{\circ}$\,--\,30.050$^{\circ}$ \\
Limiting indices  & -5\,$\leq$\,$h$\,$\leq$\,5 \\
 & -22\,$\leq$\,$k$\,$\leq$\,22 \\
 & -5\,$\leq$\,$l$\,$\leq$\,5 \\
Collected/unique reflections & 3496 / 242 \\
$R_{\rm int}$ & 0.0529 \\
Absorption correction & numerical (Gaussian) \\
Max./min. transmission & 0.194 / 0.021 \\
Data / restraints / parameters & 242 / 0 / 19 \\
Goodness of fit on $F^2$ & 1.153 \\
$R$ indices [$I\,>\,2\sigma (I)$] & $R_1$\,=\,0.0190 \\
  & $wR_2$\,=\,0.0407 \\
Extinction coefficient & 0.0015(2) \\
Largest difference peak and hole & $+$3.308 and $-$1.625\,e\AA$^{-3}$ \\
\end{tabular}
\end{ruledtabular}
\end{table}

\begin{table*}
\caption{\label{tabSI2} Refined standardized atomic coordinates, occupancy rates and isotropic displacement parameters for UNi$_{0.34}$Ge$_2$.}
\centering
\begin{ruledtabular}
\begin{tabular}{llcccll}
Atom & Wyckoff position & $x$ & $y$ & $z$ & Occupancy & $U_{\rm eq}$ (10$^{-3}$\,\AA$^2$)\\ 
\hline
U & $4c$ & 0.39618(2) & 1/4 & 0.0079(2) & 1 & 8(1) \\  
Ni & $4c$ & 0.1900(2) & 1/4 & 0.014(2) & 0.337(6) & 14(2) \\
Ge1 & $4c$ & 0.05245(9) & 1/4 & 0.0148(3) & 1 & 15(1) \\
Ge2 & $4c$ & 0.74976(9) & 1/4 & 0.0218(3) & 1 & 22(1) \\
\end{tabular}
\end{ruledtabular}
\end{table*}

\begin{figure}[t]
\centering
\includegraphics[width=0.6\columnwidth]{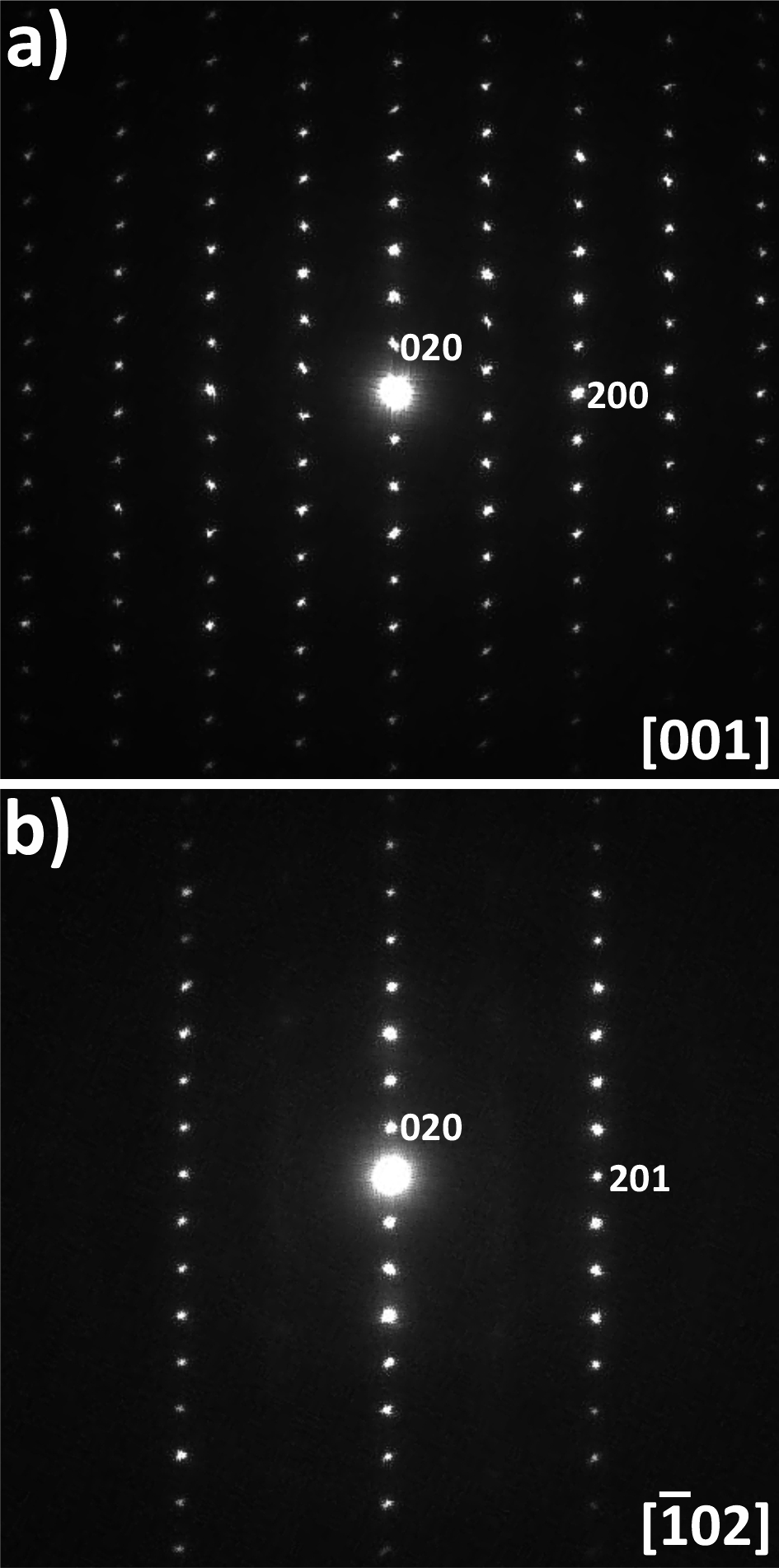}
\caption{\label{fig:tem} SAED patterns of a single crystal of UNi$_{0.34}$Ge$_2$ along two different zone axes, indexed in the orthorhombic CeNiSi$_2$-type unit cell.}
\end{figure}

The crystal structure of UNi$_{1-x}$Ge$_2$ was first checked by single crystal XRD on a small piece of the Czochralski grown sample. The structural resolution and refinements confirm the orthorhombic Ni-deficient CeNiSi$_2$-type for this phase, with cell parameters and atomic positions (see Tabs.\,\ref{tabSI1} and \ref{tabSI2}, and the .cif file in Supplemental Material) fully compatible with the previously published ones \cite{Pasturel2021}. Along the U zig-zag chains, the interatomic distances between uranium atoms are 3.8600(4)\,\AA ~and the angles 62.754(5)$^{\circ}$. These values are very close to those encountered in the U$T\!E_{1-x}$Ge$_2$ series, and slightly larger and smaller, respectively, than those found in UGe$_2$ (for a summary and discussion of these values, see Ref.\,\onlinecite{Pasturel2021}). The refined occupancy of the Ni-site leads to the chemical composition UNi$_{0.34(1)}$Ge$_2$, slightly more understoichiometric than the previously reported polycrystalline sample (UNi$_{0.45(1)}$Ge$_2$), but in good agreement with EDX data, within the 1~at.\% accuracy of the method.

To rule out the possibility for an ordered Ni-site occupancy, recently reported for different ternary rare earth germanides,\cite{Zuravleva2005,Zhang2015,Bao2021} electron diffraction has been performed. The SAED patterns along various zone axes (Fig.~\ref{fig:tem}) can be fully indexed with the UNi$_{0.34}$Ge$_2$ structure determined from single crystal XRD and no superstructure spots are visible, confirming the statistical occupancy of the Ni-site.


\subsection{\label{sec:magn-grnd-st}Magnetic ground state}

\begin{figure}
\includegraphics[width=\columnwidth]{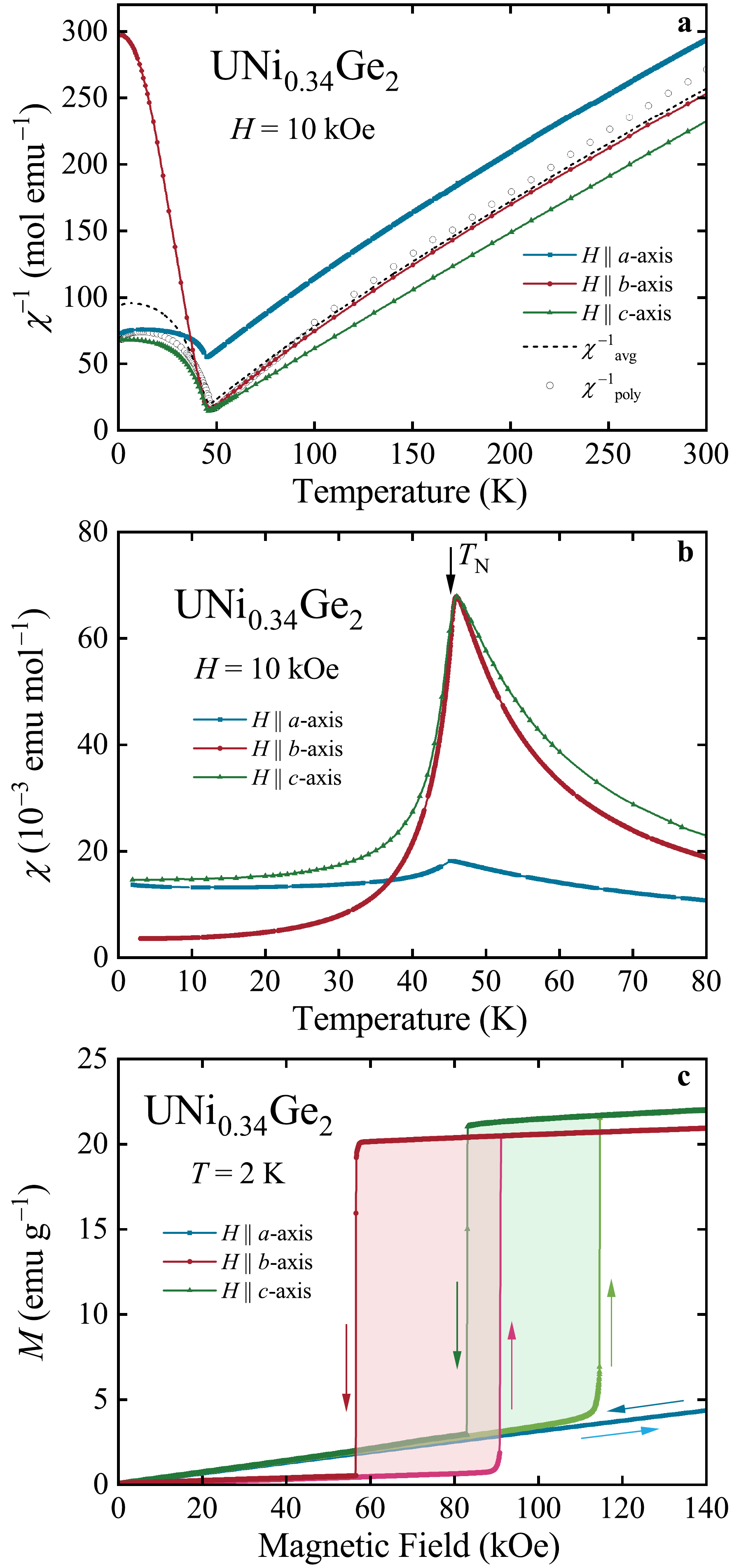}
\caption{\label{fig:magn-prop-all}(Color online) (a) Temperature dependence of inverse magnetic susceptibility $\chi^{-1}$ of single-crystalline UNi$_{0.34}$Ge$_2$ measured in magnetic field $H$ applied along the main crystallographic axes, together with its averaged values $\chi_{\rm avg}$ and the data collected earlier for polycrystalline UNi$_{0.45}$Ge$_2$ \cite{Pasturel2021}. (b) Low-temperature $\chi(T)$ of the crystal studied; an arrow indicates the N{\'e}el temperature. (c) Field variation of the magnetization $M$ measured for the main crystallographic axes with increasing and decreasing $H$, as indicated by arrows.}
\end{figure}

\begin{figure}
\includegraphics[width=\columnwidth]{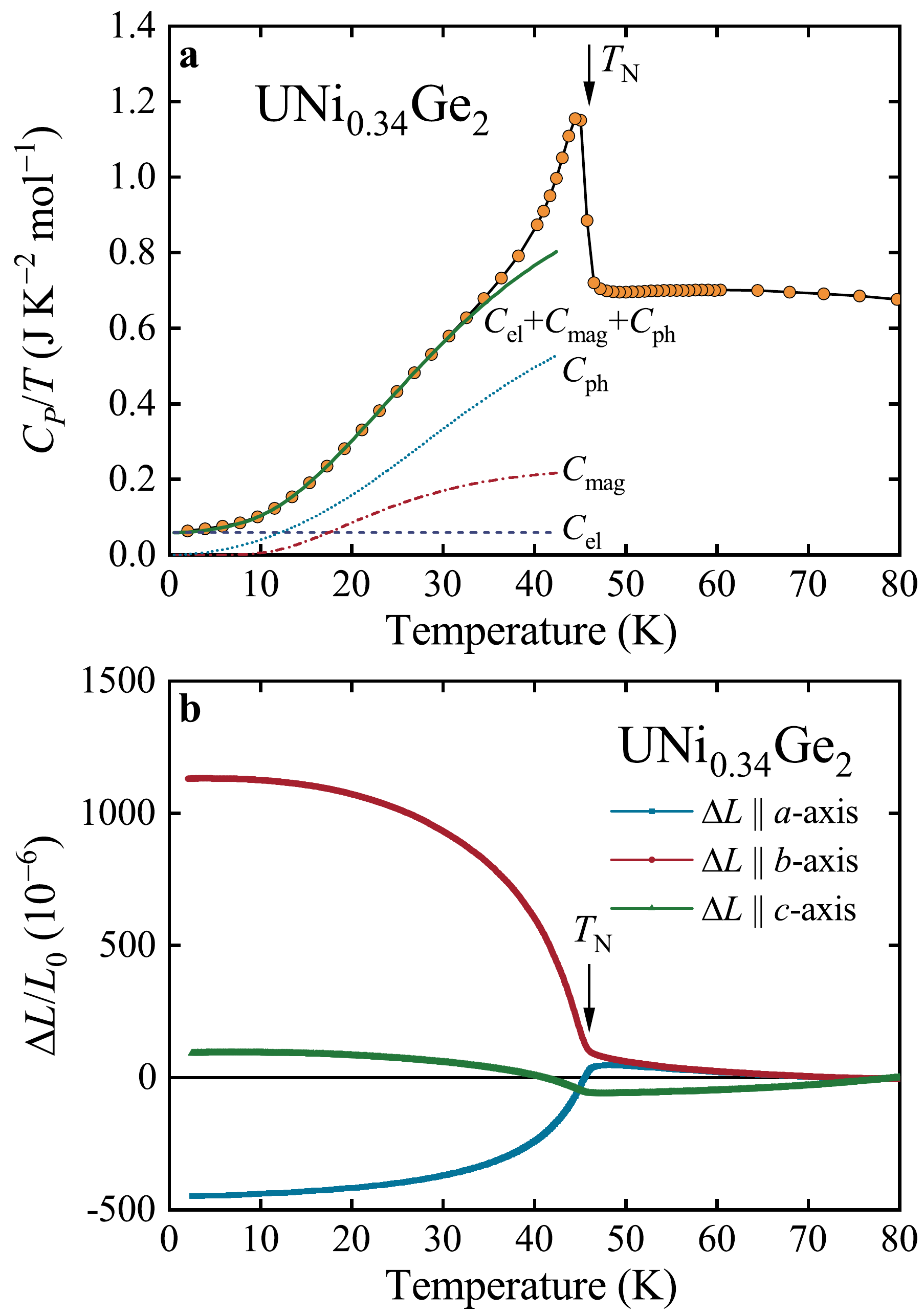}
\caption{\label{fig:therm-prop-all}(Color online) (a) Low-temperature specific heat $C_P$ of single-crystalline UNi$_{0.34}$Ge$_2$ measured in zero magnetic field and divided by $T$; an arrow marks the ordering temperature $T_{\rm N}$, and $C_{\rm el}$, $C_{\rm ph}$ and $C_{\rm mag}$ show -- respectively -- electron, phonon and magnon contributions to the total $C_P/T$ obtained from least-squares fits (for details see the text). (b) Zero-field and low-temperature linear fractional thermal expansion $\Delta L/L_0$ of the studied crystal measured along the main crystallographic directions, with $L_0$ taken at 80\,K.}
\end{figure}

\begin{table}[t]
\caption{\label{tab:CW-fits} Least-squares fitting parameters of the modified Curie-Weiss law to the experimental $\chi^{-1}(T)$ curves obtained for single-crystalline UNi$_{0.34}$Ge$_2$, plotted in Fig.\,\ref{fig:magn-prop-all}(a), compared with its averaged values (avg) and those obtained for polycrystalline UNi$_{0.45}$Ge$_2$ (poly).}
\begin{ruledtabular}
\begin{tabular}{cccc}
Crystal orientation &
$\mu_{\rm eff}$ &
$\theta_{\rm p}$ &
$\chi_0$ \\
in magnetic field &
($\mu_{\rm B}$)&
(K)&
(emu\,mol$^{-1}$)\\
\colrule
$H\parallel a$ & $2.61(1)$ & $-3.8(1)$ & $6.1(1)\times 10^{-4}$ \\
$H\parallel b$ & $2.65(1)$ & $31.4(1)$ & $6.8(2)\times 10^{-4}$ \\
$H\parallel c$ & $2.95(1)$ & $31.7(1)$ & $2.5(1)\times 10^{-4}$ \\
\colrule
avg & $2.74(1)$ & $19.8(1)$ & $5.1(1)\times 10^{-4}$ \\
poly\footnote{Ref.\,\cite{Pasturel2021}} & $2.74(1)$ & $22(4)$ & $2.9(6)\times 10^{-4}$ \\
\end{tabular}
\end{ruledtabular}
\end{table}

Fig.\,\ref{fig:magn-prop-all} displays results of the basic magnetic characterization of the single crystal of UNi$_{0.34}$Ge$_2$. As can be seen, the inverse magnetic susceptibility $\chi^{-1}$ of the compound is strongly anisotropic over the entire temperature range studied, while exhibiting above about 50\,K qualitatively the same paramagnetic Curie-like behavior for each of the main crystallographic axes. Least-squares fits of the modified Curie-Weiss law:
\begin{equation}
\chi(T) = \frac{C}{T-\theta_{\rm p}} + \chi_0
\end{equation}
to the experimental data collected for individual axes (for the sake of clarity, the fits are not shown here) yielded the effective magnetic moments $\mu_{\rm eff}$, the paramagnetic Curie-Weiss temperature $\theta_{\rm p}$, and the temperature independent Pauli-like susceptibility $\chi_0$ collected in Tab.\,\ref{tab:CW-fits}.

To compare the obtained results with the previously published data for the polycrystalline compound UNi$_{0.45}$Ge$_2$ \cite{Pasturel2021} (see open circles in Fig.\,\ref{fig:magn-prop-all}(a)), we plotted temperature dependence of the averaged magnetic susceptibility $\chi^{-1}_{\rm avg}(T)$ of the single-crystalline UNi$_{0.34}$Ge$_2$ compound, defined as $\chi_{\rm avg} (T) = \frac{1}{3}\chi_{a}(T) + \frac{1}{3}\chi_{b}(T) + \frac{1}{3}\chi_{c}(T)$, where $\chi_{a}(T)$, $\chi_{b}(T)$ and $\chi_{c}(T)$ denote the experimental curves measured for $H\parallel a$, $H\parallel b$ and $H\parallel c$, respectively (see black dashed line in Fig.\,\ref{fig:magn-prop-all}(a)). In addition, we also calculated the averaged values of the Curie-Weiss fitting parameters and included them in Tab.\,\ref{tab:CW-fits} along with the corresponding values found for the polycrystalline sample. As can be seen, both the plotted curves and the values of the corresponding fitting parameters are very close to each other, confirming the correct orientation of the studied crystal and good reproducibility of the magnetic properties of the UNi$_{1-x}$Ge$_2$ system.

As it was found in polycrystalline UNi$_{0.45}$Ge$_2$, the estimated values of the effective magnetic moment in single-crystalline UNi$_{0.34}$Ge$_2$ are much smaller than the values predicted for free U$^{3+}$ and U$^{4+}$ ions (3.62 and 3.58 $\mu_B$, respectively), but they are large enough to attribute the observed magnetic properties of the compound exclusively to the uranium ions. And the observed discrepancy is due to partial delocalization of 5f electrons, magnetocrystalline anisotropy, and/or crystal field effect.

Interestingly, the value of $\theta_{\rm p}$ derived for the single crystal of UNi$_{0.34}$Ge$_2$ is positive for the $b$ and $c$ axis (and comparable to $T_{\rm N}$), while it is negative for the $a$ axes. That hints at predominance of ferromagnetic exchange interactions between the magnetic moments of uranium along the $b$ and $c$ axes, and at antiferromagnetic coupling between them along the hard $a$ axis of the studied compound.

As can be inferred from Fig.\,\ref{fig:magn-prop-all}(b), the single-crystalline UNi$_{0.34}$Ge$_2$ compound orders antiferromagnetically below the N{\'e}el temperature $T_{\rm N}$\,=\,45.5(5)\,K (defined as a maximum in ${\rm d}\chi/{\rm d}T$, \textit{cf.} Fig.\,\ref{fig:MvsT-all}), which is slightly lower than $T_{\rm N}$\,=\,47\,K found in polycrystalline UNi$_{0.45}$Ge$_2$ and determined in lower field of 100 Oe \cite{Pasturel2021}. That small difference in the ordering temperature is fully consistent with the previously observed small decrease of $T_{\rm N}$ with decreasing Ni-content in the UNi$_{1-x}$Ge$_2$ system (for details see the Introduction in Ref.\,\onlinecite{Pasturel2021}).

Field variation of the magnetization $M$ of UNi$_{0.34}$Ge$_2$, measured in magnetic field applied along its main crystallographic axes, is shown in Fig.\,\ref{fig:magn-prop-all}(c). The $M(H)$ curve obtained for $H\parallel a$-axis is featureless: it is linear in $H$ over the entire magnetic field range studied, as expected for an antiferromagnet. At the highest applied field (\textit{i.e.} 140\,kOe), it reaches a value of 4.35(1)\,emu\,g$^{-1}$, which corresponds to the magnetic moment projection of about 0.32(1)\,$\mu_{\rm B}$, which is (as expected) far from the ordered magnetic moment of 1.95(3)\,$\mu_{\rm B}$ estimated from the neutron diffraction experiments \cite{Pasturel2021}.

If, on the other hand, $H$ is applied along the $b$ axis, then the magnetization of the studied crystal is linear in field only up to 90\,kOe, when $M$ rapidly increases (more than four times), manifesting a metamagnetic phase transition from the antiferromagnetism into the polarized paramagnetism. Above the transition, the magnetization shows a clear trend toward saturation and reaches at 140\,kOe a value of 20.93(1)\,emu\,g$^{-1}$. It corresponds to a value of the ordered magnetic moment $\mu_{\rm ord}$ of about 1.52(1)\,$\mu_{\rm B}$, which is not so far from the value of $\mu_{\rm ord}$ derived from the neutronographic data (1.95(3)\,$\mu_{\rm B}$ \cite{Pasturel2021}), and about the size expected for depopulated ground multiplet of uranium ions. When the magnetic field is lowered, the magnetization decreases to its initial values only at about 57\,kOe, exhibiting quite large magnetic hysteresis in $M(H)$ with a width of nearly 40\,kOe. The observed hysteresis along with the sharpness of the transition clearly indicates that the transition is of the first order.

$M(H)$ measured for $H\parallel c$ shows features of both the $a$ and $b$ axes. In particular, when the field is increased to about 100\,kOe, the $M(H)$ curves measured for $H\parallel c$ and $H\parallel a$ overlap. However, applying higher fields triggers for the $c$ axis the metamagnetic transition observed for the $b$ axis but not observed for the $a$ axis. Moreover, the transition found for $H\parallel c$ has similar features as that found for $H\parallel b$: it is associated with pronounced magnetic hysteresis of the width of about 31\,kOe and the saturation (at 140\,kOe) at a value of 21.96(1)\,emu\,g$^{-1}$. It corresponds to a value of the ordered magnetic moment of about 1.59(1)\,$\mu_{\rm B}$, which is as close to the neutron-diffraction value of 1.95(3)\,$\mu_{\rm B}$ as that found for $H \parallel b$. The averaged value of $\mu_{\rm ord}$ is 1.14(3)\,$\mu_{\rm B}$, which is close to the value reported for the polycrystalline sample (\textit{i.e.} 0.9\,$\mu_{\rm B}$ \cite{Pasturel2021}).

Based on the described behavior of the field dependence of the magnetization of single-crystalline UNi$_{0.34}$Ge$_2$, it can be concluded that $a$ is the hard magnetization axis of the studied crystal, while $b$ and $c$ are its easy magnetization axes. Moreover, the shapes of the $M(H)$ curves in Fig.\,\ref{fig:magn-prop-all}(c) fully agree with the positive signs of the paramagnetic Curie-Weiss temperature found for the $b$ and $c$ axes, and with negative $\theta_{\rm p}$ derived for the $a$ axis. This is a somewhat surprising finding, given the zero-field neutron diffraction data, which showed ferromagnetically ordered $ac$ planes stacked antiferromagnetically along the $b$ axis in a sequence $-++-$ and with the magnetic moments aligned along the $b$ axis \cite{Pasturel2021}. It turns out, that although the $b$ axis is clearly favored in zero field (in terms of the magnetic order), the in-plane ($ac$) interactions are far from symmetric, which is revealed only by applying magnetic field.

Fig.\,\ref{fig:therm-prop-all} shows temperature variation of other thermodynamic properties of UNi$_{0.34}$Ge$_2$, namely specific heat (plotted here as $C_P/T$) and linear fractional thermal expansion (defined as $\Delta L /L_0$, where  $\Delta L$ is the sample length change, and $L_0$ is its original length at a given temperature -- here: 80\,K). In $C_P(T)/T$ (Fig.\,\ref{fig:therm-prop-all}(a)), the antiferromagnetic phase transition manifests itself as a distinct $\lambda$-shaped peak occurring just below $T_{\rm N}$ (determined from the magnetic properties data), confirming the intrinsic character of the ordering and high quality of the crystal, and being in excellent agreement with the data reported for the polycrystalline sample. As in the case of polycrystalline sample, we analysed $C_P(T)$ of the single crystal in terms of a sum of electron, phonon and magnon contributions, \textit{i.e.}:
\begin{equation}
\label{eq:specific-heat}
C_P (T) = C_{\rm el}(T) + C_{\rm mag}(T) + C_{\rm ph}(T),
\end{equation}
where
\begin{equation}
C_{\rm el}(T) = \gamma T
\end{equation}
describes the conduction-band electron contribution with the Sommerfeld coefficient $\gamma$,
\begin{equation}
C_{\rm mag}(T) = \alpha T^{-0.5} e^{-\Delta/T} 
\end{equation}
represents the antiferromagnetic magnon contribution according to Ref.\,\cite{Akhiezer1961} with the coefficient $\alpha$ and the energy gap $\Delta$ in the spin-waves spectrum, and
\begin{equation}
C_{\rm ph}(T) = 9rR\left( \frac{T}{\Theta_{\rm D}} \right)^3 \int^{\Theta_{\rm D}/T}_0 \frac{x^4 e^x}{\left( e^x -1\right)^2}
\end{equation}
is the phonon contribution in a form of the conventional Debye equation, with $r$ as the number of atoms in the formula unit, the universal gas constant $R$, and the characteristic Debye temperature $\Theta_{\rm D}$. The least-squares fitting of Eq.\,(\ref{eq:specific-heat}) to the experimental data below about 30\,K yielded for single-crystalline UNi$_{0.34}$Ge$_2$ the parameters: $\gamma$\,=\,59(1)\,mJ\,K$^{-2}$mol$^{-1}$, $\Theta_{\rm D}$\,=\,255(2)\,K, $\alpha$\,=\,376(9)\,J\,K$^{-0.5}$\,mol$^{-1}$, and $\Delta$\,=\,78.0(4)\,K, which are in satisfactory agreement with those obtained for polycrystalline UNi$_{0.45}$Ge$_2$, \textit{i.e.} $\gamma$\,=\,54.2(6)\,mJ\,K$^{-2}$mol$^{-1}$, $\Theta_{\rm D}$\,=\,267(2)\,K, $\alpha$\,=\,486(11)\,J\,K$^{-0.5}$\,mol$^{-1}$, and $\Delta$\,=\,78.7(2)\,K \cite{Pasturel2021}.

It is worth noting that the proposed model is very simplified when it comes to the description of magnon and phonon spectra, and electron contributions. Therefore, the obtained values of the fitting parameters should be considered only as a rough estimate of the corresponding energy scales. Nevertheless, even such a fit gives some estimation of the phonon contribution, which can be used to calculate the non-phonon entropy at $T_{\rm N}$. One can easily show (see Fig.SM1 in Supplementary Materials) that the non-phonon entropy in UNi$_{0.34}$Ge$_2$ reaches at $T_{\rm N}$ a value of about 9.1(2)\,J\,K$^{-1}$\,mol$^{-1}$, which is very close to $R\ln{3}$ (9.13 \,J\,K$^{-1}$\,mol$^{-1}$) expected for 3 populated levels of the uranium ground multiplet split in the crystal field –- the situation very likely in the studied compound and at the considered temperatures.

As can be seen in Fig.\,\ref{fig:therm-prop-all}(b), the thermal expansion of the compound studied, measured in zero magnetic field along its main crystallographic axes, is weakly temperature-dependent in the range from 80\,K down to the ordering temperature $T_{\rm N}$, at which rapid changes in the lattice dimensions occur, characteristic of the invar effect \cite{Guillaume1897}. In particular, the unit cell of the studied compound expands along the $b$ and $c$ axes by $+$1130(1)\,ppm and 90(1)\,ppm, respectively, while it shrinks along the $a$ axis by $-$423(1)\,ppm. The expansion of the crystal lattice along the easy magnetic direction and its shrinking along the hard axis implies a strong coupling between spins and lattice in the antiferromagnetic state \cite{Klimczuk2012, Guillaume1897}. Furthermore, the magnitude of that effect in zero field is very large, compared to other uranium compounds studied so far, \textit{e.g.} UO$_2$ (about 40\,ppm \cite{Jaime2017}) and UN (about 250\,ppm \cite{Shrestha2017}). Deeper analysis of that property of UNi$_{0.34}$Ge$_2$ is in progress and will be published later on.


\subsection{\label{sec:metamagn-trans} Metamagnetic phase transition}

\begin{figure}
\includegraphics[width=\columnwidth]{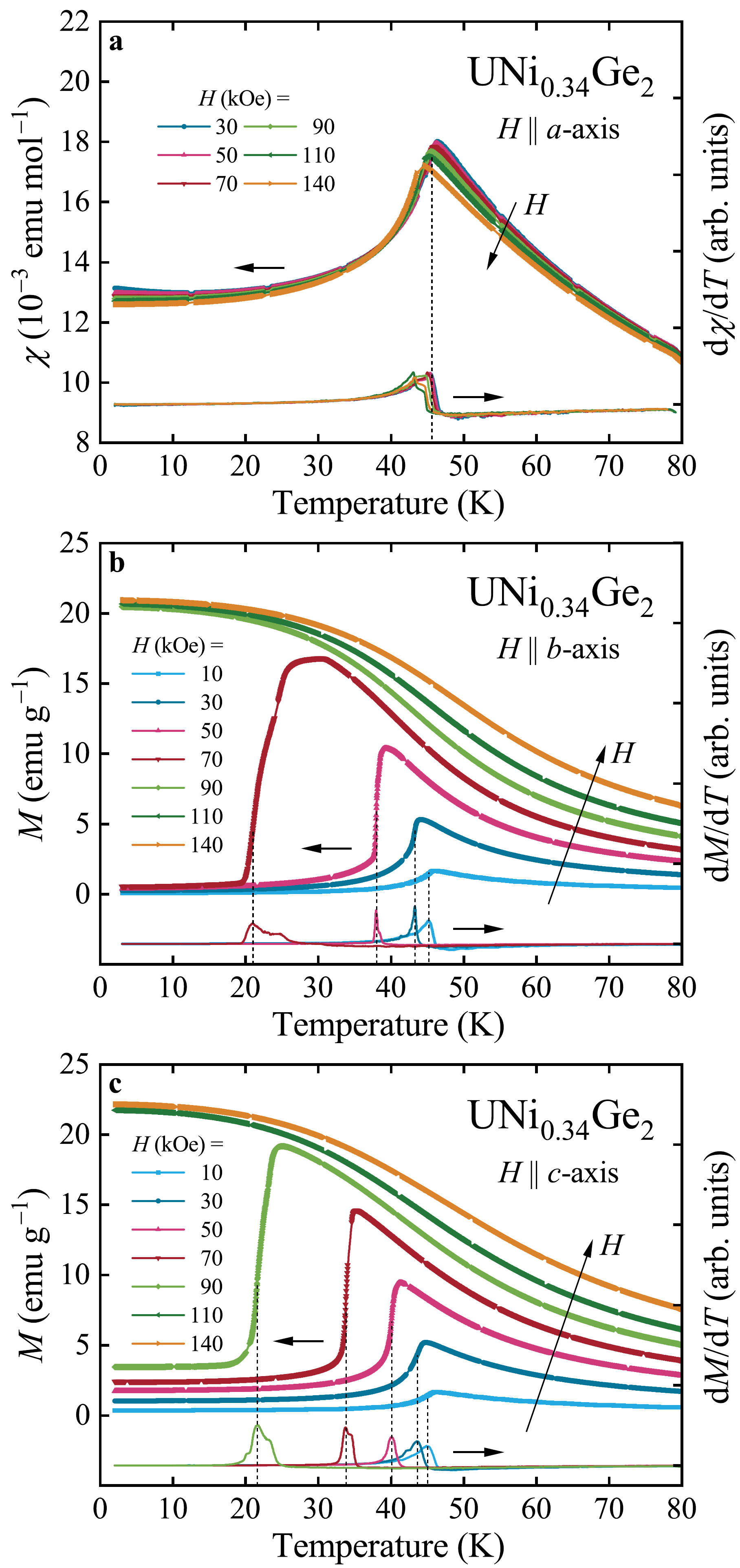}
\caption{\label{fig:MvsT-all}(Color online) Left axes: dc magnetic susceptibility $\chi$ and magnetization $M$ of single-crystalline UNi$_{0.34}$Ge$_2$ measured as a function of decreasing temperature in various, constant applied magnetic fields; diagonal arrows indicate the increase of the magnetic field $H$. Right axes: temperature derivatives of the experimental curves; dashed vertical lines points at deduced transition temperatures (for clarity, only one dashed line is shown in panel (a)).}
\end{figure}

\begin{figure*}
\centering
\includegraphics[width=2\columnwidth]{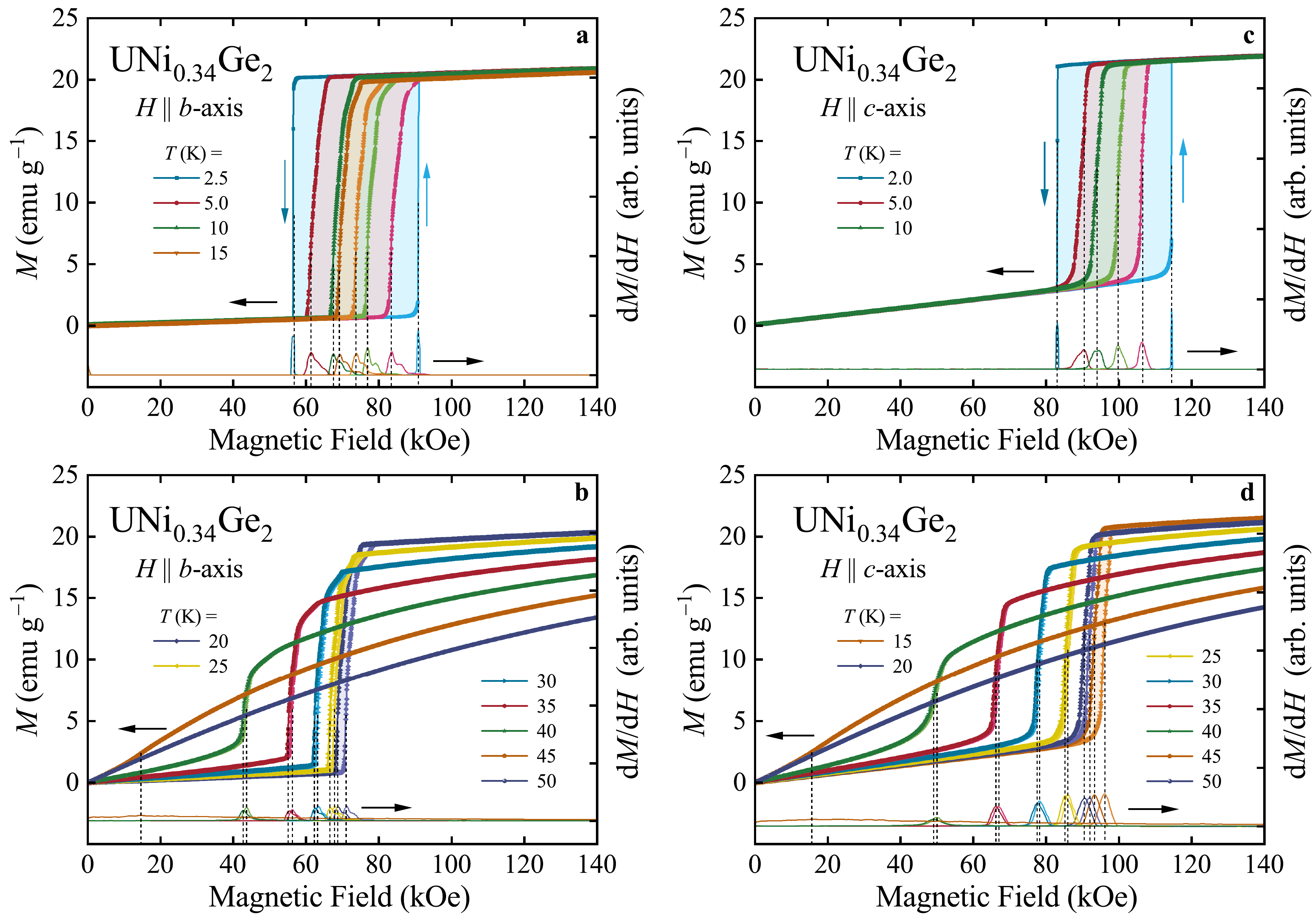}
\caption{\label{fig:MvsH-b-c}(Color online) Magnetization of single-crystalline UNi$_{0.34}$Ge$_2$ measured at various temperatures as a function of (a,b) $H \parallel b$ and (c,d) $H \parallel c$; vertical arrows indicate field changes for the largest hysteresis (the arrows for other isotherms are omitted for clarity). Right axes: field derivatives of $M(H)$; dashed lines indicate positions of maxima in ${\rm d}M/{\rm d}H$.}
\end{figure*}

\begin{figure}
\centering
\includegraphics[width=\columnwidth]{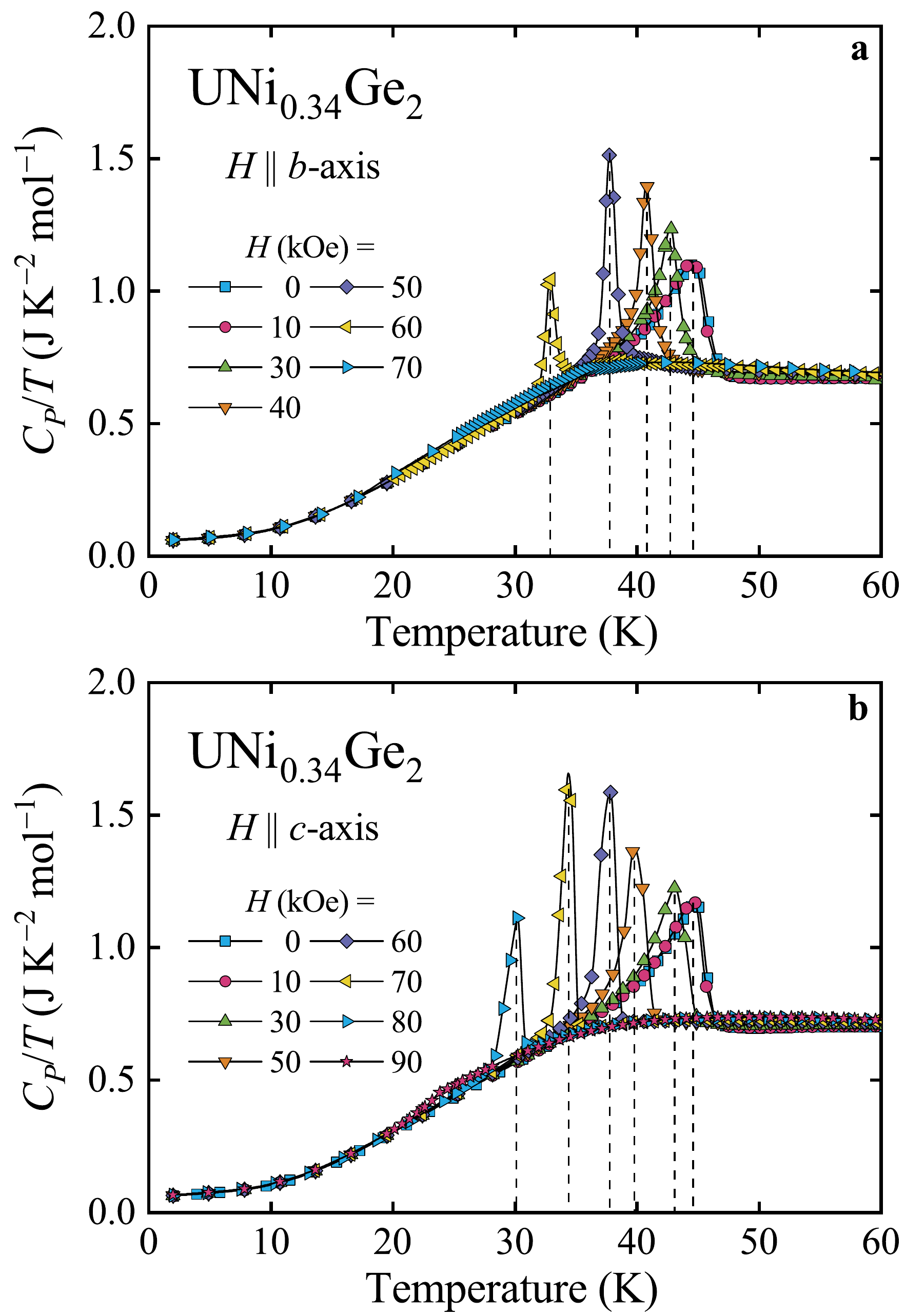}
\caption{\label{fig:CvsT-bc}(Color online) Temperature dependence of specific heat $C_P$ of single-crystalline UNi$_{0.34}$Ge$_2$ measured with decreasing $T$ in various magnetic fields applied  parallel to the $b$- (a) and $c$-axis (b); dashed lines indicate positions of maxima in $C_P (T)/T$.}
\end{figure}

\begin{figure}
\centering 
\includegraphics[width=\columnwidth]{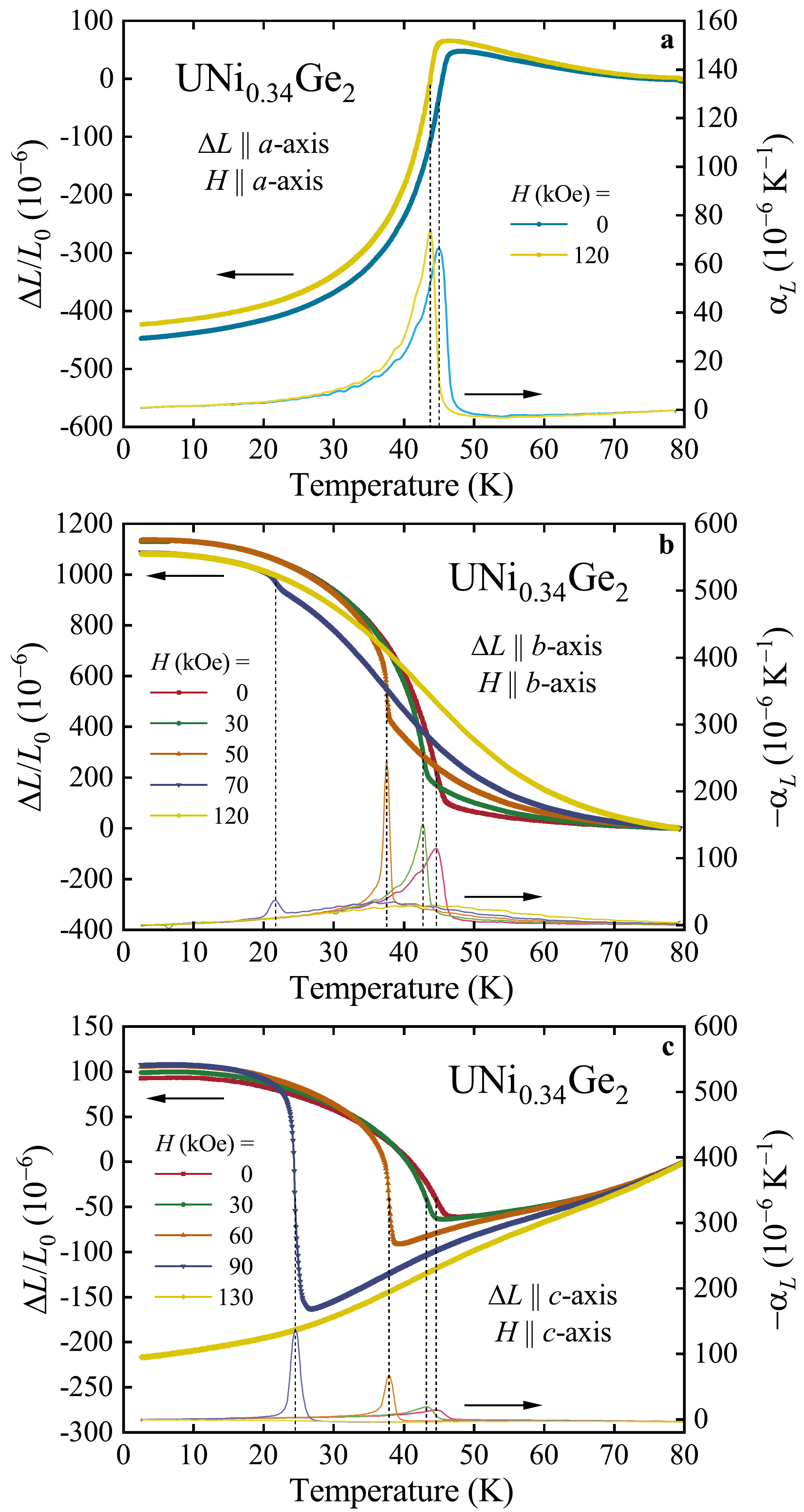}
\caption{\label{fig:LvsT-all}(Color online) Temperature dependence of linear thermal expansion $\Delta L /L_0$ (left axes) and linear thermal expansion coefficient $\alpha_L$  (right axes) of single-crystalline UNi$_{0.34}$Ge$_2$ measured with deacreasing $T$ along the main crystallographic axes in longitudinal magnetic field; dashed lines indicate phase transitions temperatures.}
\end{figure}

\begin{figure*}
\includegraphics[width=2\columnwidth]{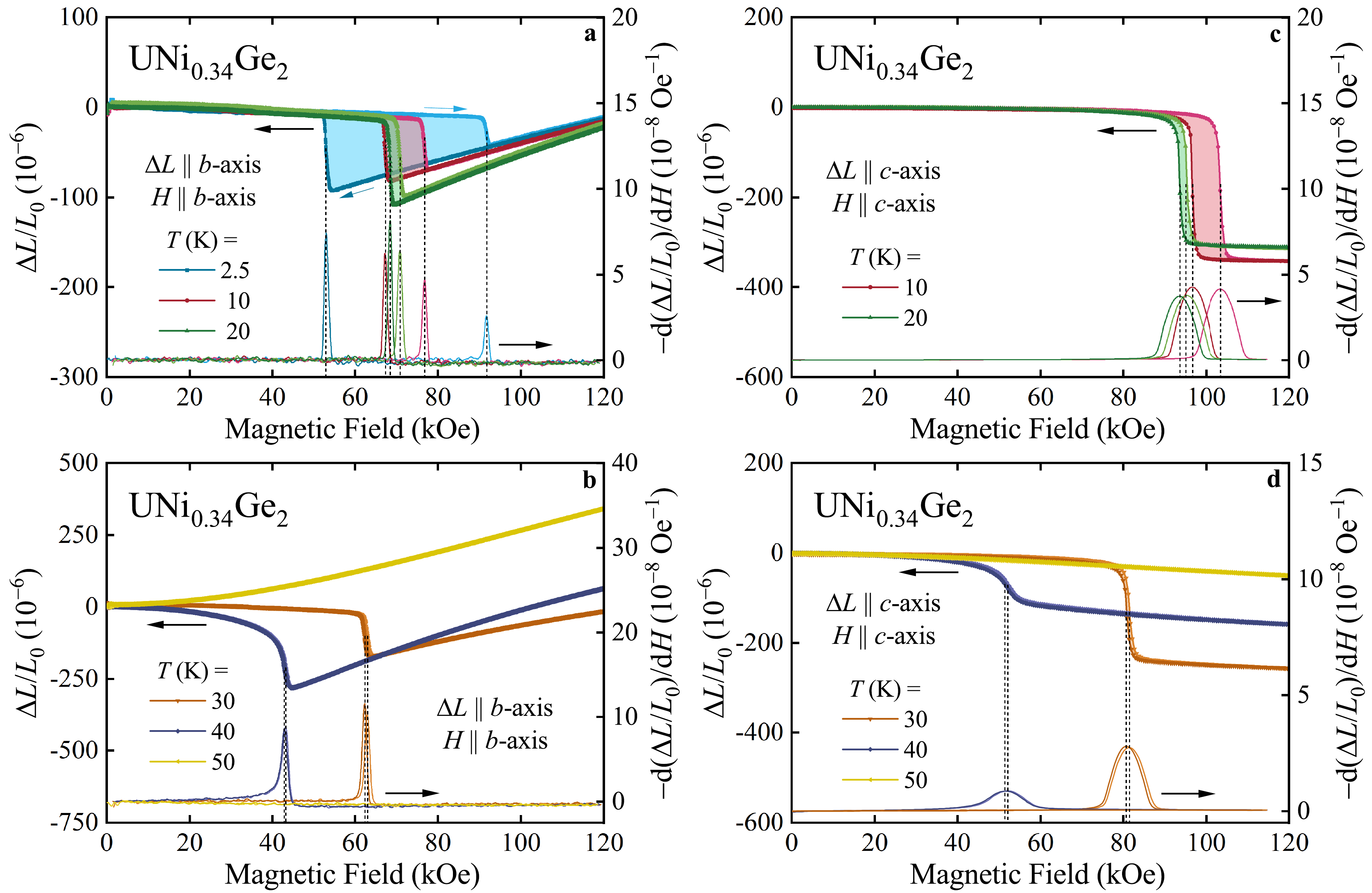}
\caption{\label{fig:LvsH-b-c}(Color online) Left axes: linear thermal expansion $\Delta L /L_0$ of single-crystalline UNi$_{0.34}$Ge$_2$ measured along the $b$ axis (panels a and b) and along the $c$ axis of the crystal (panels c and d) as a function of increasing and decreasing longitudinal magnetic field; sloped arrows indicate field changes for the largest hysteresis (the arrows for other isotherms are omitted for clarity). Right axes: field derivatives of $\Delta L /L_0$.}
\end{figure*}

To shed more light on the nature of the metamagnetic phase transition evidenced in the studied compound and on its unexpected occurrence in magnetic fields applied along the $c$-axis of the crystal, we extended our measurements of thermodynamic properties of UNi$_{0.34}$Ge$_2$ to higher fields. Fig.\,\ref{fig:MvsT-all} shows temperature dependence of its magnetization (for convenience, it is plotted in panel (a) as dc magnetic susceptibility), measured in various constant magnetic fields. As can be seen, the magnetic field applied along the $a$ axis, which turned out to be the hard axis of the crystal, has virtually no effect on the shape and position of the antiferromagnetic phase transition at least up to 140\,kOe. One can only observe a barely visible shift in $T_{\rm N}$ toward lower temperatures, as expected for an antiferromagnet. 

The field-induced shift of $T_{\rm N}$ toward lower temperatures is much more pronounced when $H$ is applied along the $b$ and $c$ axes, which are the easy magnetization axes. As can be seen in Figs.\,\ref{fig:MvsT-all}(b) and \ref{fig:MvsT-all}(c), the antiferromagnetic cusp in $M(T)$ rapidly shifts its position with increasing magnetic field, and the N{\'e}el temperature derived from ${\rm d}M/{\rm d}T$ (see dashed lines in Figs.\,\ref{fig:MvsT-all}(b) and \ref{fig:MvsT-all}(c), right axis) lowers from about 45\,K in 10\,kOe to about 21\,K in 70\,kOe for the $b$ axis, and from about 45\,K in 10\,kOe to about 22\,K in 90\,kOe for the $c$ axis. In higher magnetic fields, the overall shape of the $M(T)$ curves changes abruptly, taking that predicted by the mean-field model for a ferromagnetic phase transition transformed by an external magnetic field (see e.g.\,Ref.\,\cite{Blundell2001}).

The behavior of all the $M(T)$ curves shown in Fig.\,\ref{fig:MvsT-all} is fully consistent with the results of the initial characterization described in Sec.\,\ref{sec:magn-grnd-st}. The antiferromagnetic ordering temperatures, defined as a maximum in the first derivative of the magnetic susceptibility (${\rm d}\chi/{\rm d}T$) or the magnetization (${\rm d}M/{\rm d}T$), were determined for each applied field (see Fig.\,\ref{fig:MvsT-all}, right axes) and plotted as the corresponding graphs in Fig.\,\ref{fig:phase-diag}.

As expected, the $M(H)$ curves measured in $H\parallel a$ are featureless and exhibit purely linear behavior in both the paramagnetic and antiferromagnetic states, hence they are not shown here as irrelevant to the study of the metamagnetic phase transition. Fig.\,\ref{fig:MvsH-b-c} displays field dependence of the magnetization of single-crystalline UNi$_{0.34}$Ge$_2$ measured at various temperatures and in magnetic field parallel to the two easy axes, \textit{i.e.} $H\parallel b$ and $H \parallel c$.  The hysteresis associated with the field-induced metamagnetic phase transition is clearly visible in both easy axes throughout the ordered region, although its shape and position change with temperature. In particular, at the lowest temperature studied (\textit{i.e.} 2.5\,K), the magnetization jumps (observed with both increasing and decreasing fields) are very sharp, and the hysteresis loops are nearly rectangular and very wide, suggesting that the metamagnetic transition studied is a first-order spin-flip (and not spin-flop) transition in both $H\parallel b$ and $H\parallel c$. As the temperature increases, the hysteresis loops for both crystallographic directions noticeably tilt toward higher magnetic fields and become narrower, but their positions remain almost unchanged up to 15\,K in $H\parallel b$ and 10\,K in $H\parallel c$ (see Figs.\,\ref{fig:MvsH-b-c}(a) and \ref{fig:MvsH-b-c}(c)). At higher $T$, the hysteresis loops quickly become hardly visible, their position shifts toward lower magnetic fields, and the $M(H)$ curves evolve into an S-shape (see Figs.\,\ref{fig:MvsH-b-c}(b) and \ref{fig:MvsH-b-c}(d)), suggesting continuous nature of the metamagnetic phase transition at temperatures approaching $T_{\rm N}$. 

Using the first field derivatives of the magnetization, ${\rm d}M/{\rm d}H$, we derived the transition fields for each $M(H)$ isotherm (see the right axes in Figs.\,\ref{fig:MvsH-b-c}(a), \ref{fig:MvsH-b-c}(b), \ref{fig:MvsH-b-c}(c) and \ref{fig:MvsH-b-c}(d)). The so-obtained data points were added to Fig.\,\ref{fig:phase-diag}.

The field-induced change in the ordering temperature of UNi$_{0.34}$Ge$_2$ is also clearly seen in the specific heat of the compound (Fig.\,\ref{fig:CvsT-bc}) and is fully consistent with that observed in magnetic properties of the compound. In particular, as the magnetic field increases, the anomaly in $C_P(T)/T$ shifts towards lower temperatures, confirming the antiferromagnetic nature of the ordering. It also changes its form from a $\lambda$ shape to a spike-like contour, thus suggesting the evolution of the magnetic phase transition from the second to the first order. Finally, at $H>$\,70\,kOe  for $H\parallel b$ and at $H>$\,90\,kOe  for $H\parallel c$ the anomaly becomes suppressed, smeared, and hardly visible in the $C_P(T)/T$ curves. The positions of the maxima in $C_P(T)/T$ were used to construct the magnetic phase diagram of UNi$_{0.34}$Ge$_2$ (Fig.\,\ref{fig:phase-diag}).

Fig.\,\ref{fig:LvsT-all} (left axes) shows the low-temperature thermal expansion of the studied compound, measured along its main crystallographic axes in longitudinal magnetic field. As can be seen, when the field is applied along the $a$-axis, the position of the phase transition remains almost unchanged, while for $H\parallel b$ and $H\parallel c$ the transition temperature decreases from about 45\,K in zero field to about 22\,K in $H$\,=\,70\,kOe for the $b$-axis, and to about 25\,K in $H$\,=\,90\,kOe for the $c$-axis. Moreover, as can be inferred from the temperature dependence of the linear thermal expansion coefficient $\alpha_L$ (see Fig.\,\ref{fig:LvsT-all}, right axes), defined as:
\begin{equation}
\alpha_L = -\frac{1}{L_0} \left( \frac{{\rm d} L}{{\rm d} T} \right)_P,
\end{equation}
the anomaly in $\alpha_L (T)$ (manifesting the phase transition in UNi$_{0.34}$Ge$_2$) clearly changes its form from the $\lambda$ shape in zero magnetic field to the spike-like shape in high fields. Note that similar evolution (interpreted as a change from the second-order to the first-order character of the phase transition) was observed also in the specific heat of the compound (Fig.\,\ref{fig:CvsT-bc}). The positions of the maxima in $\alpha_L (T)$ were added to the magnetic phase diagram of UNi$_{0.34}$Ge$_2$ (Fig.\,\ref{fig:phase-diag}).

To verify the presence of the magnetic hysteresis observed in $M(H)$ for $H \parallel b$-axis and  $H\parallel c$-axis (Fig.\,\ref{fig:MvsH-b-c}), we measured the linear thermal expansion of single-crystalline UNi$_{0.34}$Ge$_2$ as a function of increasing and decreasing longitudinal magnetic field, for the two easy axes. As can be seen, in the magnetically ordered region $\Delta L /L_0$ (Fig.\,\ref{fig:LvsH-b-c}, left axes) exhibits distinct field-induced phase transitions for both crystal orientations, visible as sharp peaks in its field derivatives (Fig.\,\ref{fig:LvsH-b-c}, right axes).  This clearly suggests the discontinuous nature of these phase transitions, which is most pronounced at the lowest temperature studied (as the hysteresis is the widest there) and less noticeable at temperatures approaching $T_{\rm N}$, being fully consistent with the magnetic properties of the system studied. Also, the phase transitions temperature and field obtained from ${\rm d}(\Delta L /L_0)/{\rm d}H$ are in good agreement with those determined from ${\rm d}M/{\rm d}H$ (see the phase diagrams in Figs.\,\ref{fig:phase-diag}(b) and Figs.\,\ref{fig:phase-diag}(c)).

Another interesting feature of the magnetostriction of UNi$_{0.34}$Ge$_2$ is its sensitivity to the external magnetic field in different regimes. As shown in Fig.~\ref{fig:LvsH-b-c}, $\Delta L/L_0$ measured along the $b$- and $c$-axis is almost independent of the field in the antiferromagnetically ordered region, while it increases almost linearly with an increasing field in the polarized paramagnetic state. Interestingly, the field-induced sample expansion is very significant - it has a similar magnitude to its contraction caused by the metamagnetic phase transition.

At first glance, one could attribute the behavior of $\Delta L (H)/L_0$ in high fields to the increasing polarization of the uranium magnetic moments with an increasing magnetic field, as would be expected for a spin-flop transition (just above it) and in the paramagnetic region. Such an interpretation, however, contradicts the results of the magnetization measurements (Fig.~\ref{fig:MvsH-b-c}), which clearly show a saturation of $M$ already above the phase transition, thus ruling out its spin-flop nature. To elucidate the microscopic origin of the field-induced expansion of UNi$_{0.34}$Ge$_2$ along the $b$- and $c$-axes, additional studies are needed, including \textit{e.g.} high-resolution XRD at low temperature and in high magnetic field, complemented by \textit{ab-initio} electron and phonon structure calculations.


\subsection{\label{sec:phase-diag}Magnetic phase diagram}

\begin{figure}[ht]
\includegraphics[width=0.9\columnwidth]{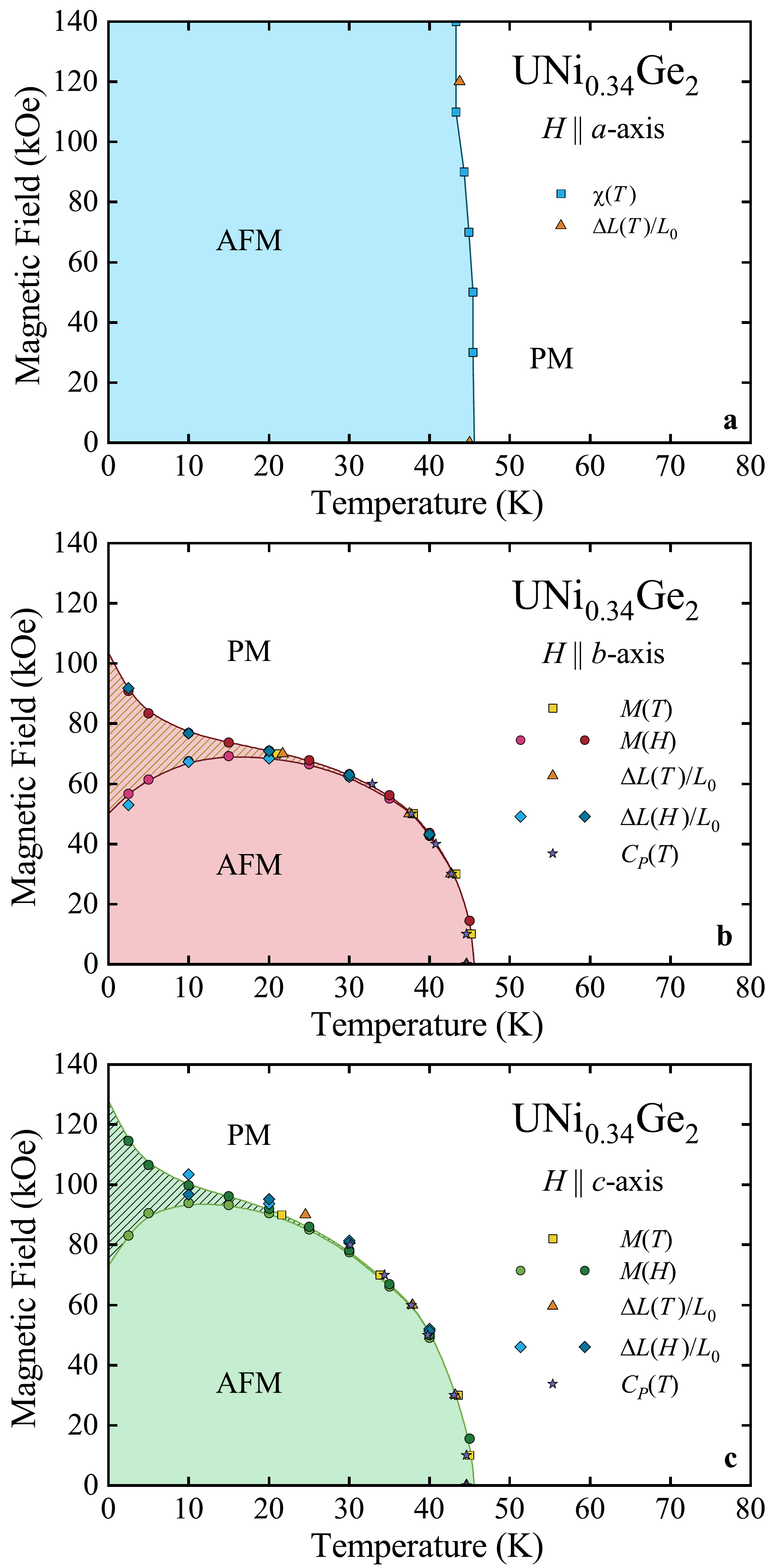}
\caption{\label{fig:phase-diag}(Color online) Tentative magnetic phase diagram constructed for single-crystalline UNi$_{0.34}$Ge$_2$ for the magnetic field directions parallel to the main crystallographic axes. AFM and PM denote an antiferromagnetic and paramagnetic state, respectively. Hatched areas indicate magnetic hysteresis associated with a metamagnetic transition.}
\end{figure}

\begin{figure}[t]
\includegraphics[width=0.9\columnwidth]{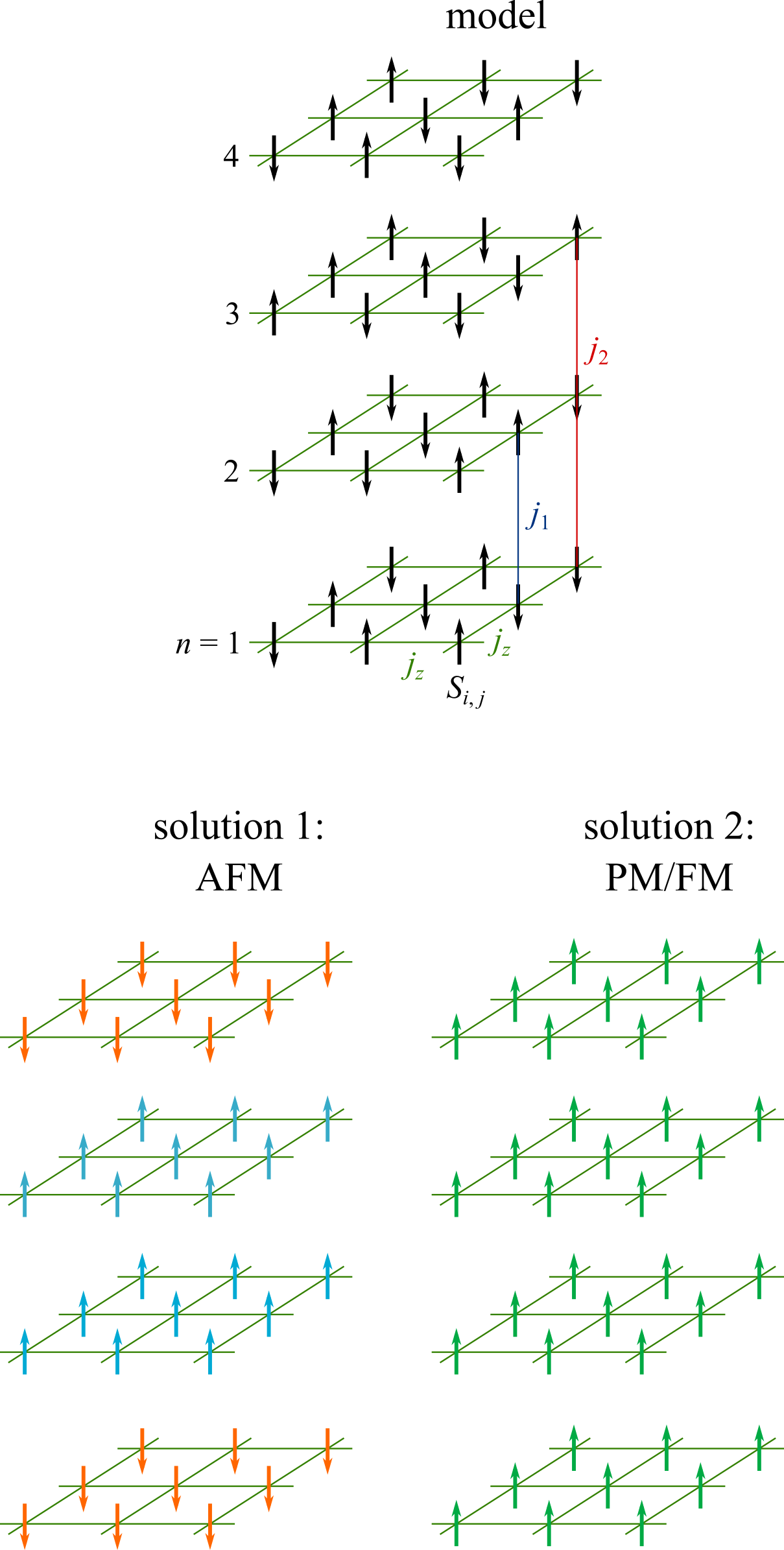}
\caption{\label{fig:model}(Color online) Four-layer Ising model used to describe the magnetic phase diagram of UNi$_{1-x}$Ge$_2$ along with its two solutions (see Sec.~\ref{sec:phase-diag} for details).}
\end{figure}

\begin{figure}[t]
\includegraphics[width=0.9\columnwidth]{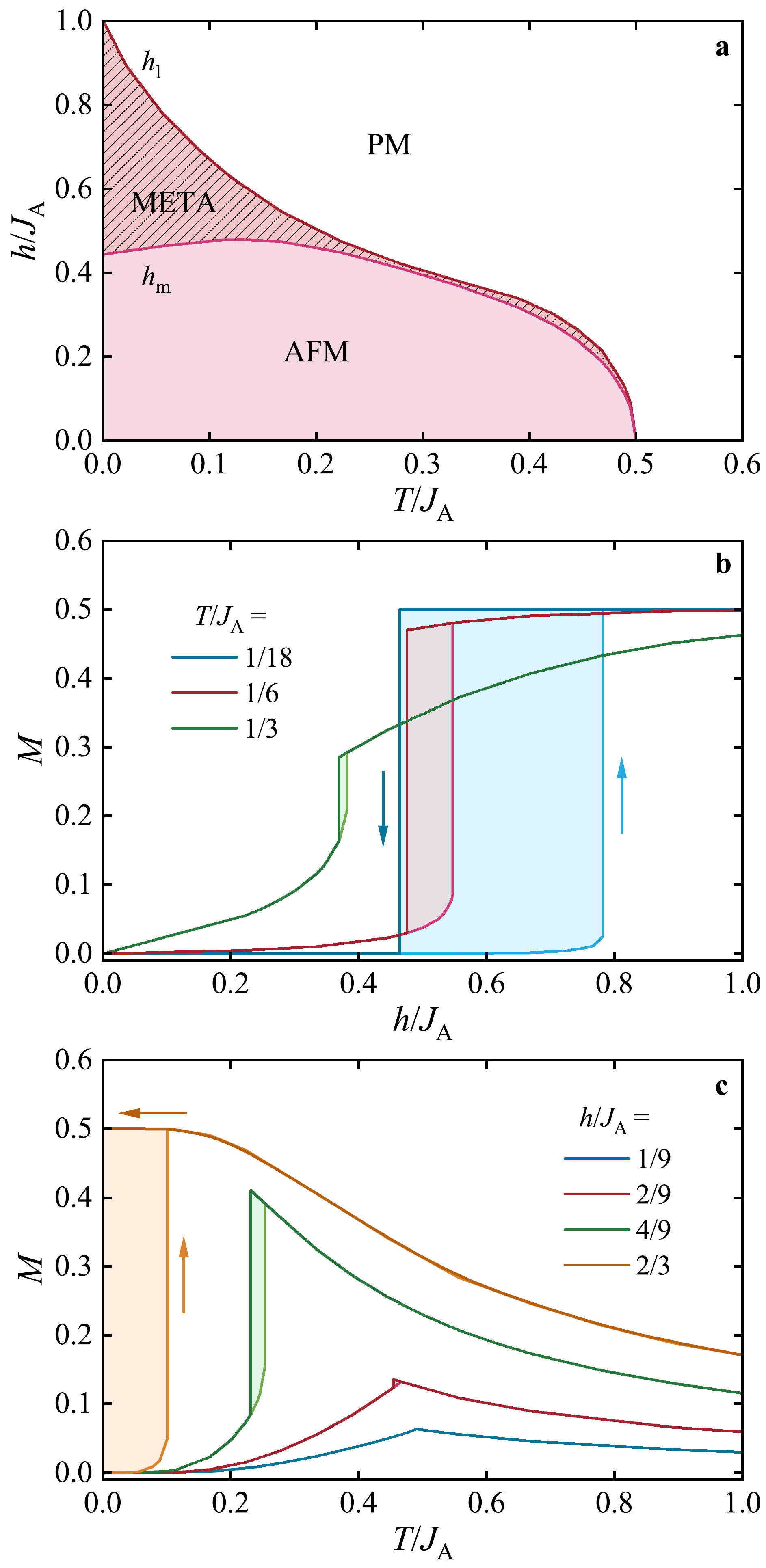}
\caption{\label{fig:theory}(Color online) (a) Theoretical magnetic phase diagram of the four-layer Ising model (for details, see the text) with three areas representing: antiferromagnetism AFM, metastable antiferromagnetism META, and paramagnetism PM; magnetic field $h$ and temperature $T$ are shown in values of $J_{\rm A}$, and the curves $h_l$ and $h_m$ represent phase boundaries. (b) Field variations of magnetization $M$ of the four layers for several values of reduced temperature $T/J_{\rm A}$; shaded areas represent the metastable state. (c) Temperature dependence of $M$ for several values of reduced magnetic field $h/J_{\rm A}$.}
\end{figure}

Fig.~\ref{fig:phase-diag} shows the tentative magnetic phase diagram of UNi$_{0.34}$Ge$_2$ constructed for each of its main crystallographic directions based on the experimental data described in Sec.~\ref{sec:metamagn-trans}. As expected, the diagram obtained for the hard axis $a$ (Fig.~\ref{fig:phase-diag}(a)) is trivial and shows only a single, antiferromagnetically ordered region with a nearly field-independent phase boundary. Since this boundary line shows no tendency to bend toward lower temperatures in higher magnetic fields, one could argue that inducing a metamagnetic transition along the hard axis would require magnetic field order of magnitude higher than that used in our experiments.

This would not be the first case of such strong magnetic anisotropy. As an extreme example of the experimental realization of such a situation, it is worth recalling the compound UN, which needed a magnetic field as high as 58~T to induce the reversal of only 1 of the 12 antiferromagnetically coupled ferromagnetic layers in the direction of the magnetic field \cite{Troc2016,Shrestha2017}. In addition, it was estimated that any further metamagnetic phase transitions in this compound would require increasingly higher magnetic fields, which is beyond experimental capability. Presumably, fully polarized paramagnetism in UN can only be observed in a field of about 258~T \cite{Troc2016}.

In the case of the magnetic field applied along the two easy axes ($b$ and $c$) the situation is quite different. The metamagnetic phase transition in UNi$_{0.34}$Ge$_2$ is observed for both orientations in moderately high magnetic fields. The average critical field, extrapolated to the absolute zero temperature, appears to be no higher than about 70-75 kOe for $H\parallel b$-axis (Fig.~\ref{fig:phase-diag}(b)) and about 95-100 kOe for $H\parallel c$-axis (Fig.~\ref{fig:phase-diag}(c)). However, the hysteresis observed in the field dependence of the magnetization and thermal expansion (Figs.~\ref{fig:MvsH-b-c} and \ref{fig:LvsH-b-c}) forms a distinct metastable region in the $H-T$ diagrams, whose width at absolute zero temperature can be estimated to be as high as about 53~kOe and 55~kOe for the $b$ and $c$ axes, respectively. This metastable region becomes increasingly narrow as the temperature increases and is hardly noticeable above about 30~K for $H\parallel b$-axis and $H\parallel c$-axis.

To reproduce the experimental magnetic phase diagram constructed for the $b$-axis of the single crystalline UNi$_{0.34}$Ge$_2$ (Fig.~\ref{fig:phase-diag}(b)), and field and temperature dependence of its magnetization (Figs.~\ref{fig:MvsH-b-c}(a,b) and Fig.~\ref{fig:MvsT-all}(b), respectively) we considered a minimal, multi-layer Ising spins model (Fig.~\ref{fig:model}; upper panel). The Hamiltonian of such a system has the form:
\begin{eqnarray}
\label{eq:hamiltonian}
H &=& -\sum_n \Bigg[ j_z \sum_{i,j} S^n_i S^n_j \nonumber \\
&+& \sum_{i} \Big( j_1 S^n_i  S^{n+1}_i + j_2 S^n_i  S^{n+2}_i \Big)  + h \sum_i S_i^n \Bigg],
\end{eqnarray}
where $S_i^n$ denotes the $z$-th component of the spin $\frac{1}{2}$, $i$ numbers the spins in the plane, and $n$ numbers the planes. The parameter $j_z$ describes the interaction between the spins in a plane, $j_1$ -- the interaction between the nearest planes, $j_2$ --  the interaction between the next-nearest planes, and $h$ stands for the external magnetic field. To describe the actual magnetic structure of UNi$_{0.34}$Ge$_2$ \cite{Pasturel2021}, we consider four layers with magnetization of individual layers $m_n= \langle S^n \rangle$ ($n$~=~1, 2, 3 and 4).

In the simplest approach, \textit{i.e.} the mean field approximation (MFA), it is easy to find the free energy $F$ in the form:
\begin{eqnarray}
\label{eq:freeenergy}
F &=& j_z (m_1^2+m_2^2+m_3^2+m_4^2) \nonumber \\
&+&  j_1(m_1+m_3)(m_2+m_4) + 2 j_2(m_1 m_3+m_2 m_4) \nonumber \\
&-& T \log \left[ 2 \cosh \frac {h+2 j_z m_1+j_1(m_2+m_4)+2 j_2 m_3}{2 T} \right ] \nonumber \\
&-& T \log \left[ 2 \cosh \frac {h+2 j_2 m_1+j_1(m_2+m_4)+2 j_z m_3}{2 T} \right ] \nonumber \\
&-& T \log \left[ 2 \cosh \frac {h+2 j_z m_2+j_1(m_1+m_3)+2 j_2 m_4}{2 T} \right ] \nonumber \\
&-& T \log \left[ 2 \cosh \frac {h+2 j_2 m_2+j_1(m_1+m_3)+2 j_z m_4}{2 T} \right ], \nonumber \\
\end{eqnarray}
and  the necessary conditions for the minimum of $F$:
\begin{equation}
\frac{\partial F}{\partial m_n} = 0. 
\end{equation}
Assuming the solutions in the form (\textit{cf.} the magnetic structure from Ref.~\onlinecite{Pasturel2021}):
\begin{equation}
m_4 = m_1  \quad  \text{and}  \quad m_3 = m_2
\end{equation}
with the notation: 
\begin{equation}
m_1 = -m+q \quad \text{and}  \quad m_2=m+q,
\end{equation}
one can find the condition for the free energy minimum from the system of equations:
\begin{eqnarray}
\label{eq:set-eqns}
4(J_F q+J_{\rm A} m)-(J_{\rm F}-J_{\rm A})t_m-(J_{\rm F}+J_{\rm A}) t_p&=&0,  \nonumber \\
4(J_F q-J_{\rm A} m)-(J_{\rm F}+J_{\rm A})t_m-(J_{\rm F}-J_{\rm A}) t_p&=&0, \nonumber \\  
\end{eqnarray}
where 
\begin{equation}
J_{\rm F} =j_z+ j_1 +j_2, \quad J_{\rm A} = j_z - j_2,
\end{equation}
and
\begin{equation}
\begin{aligned}
t_p = \tanh \left( \frac{h+2 J_{\rm A} m+2 J_{\rm F} q}{2 T} \right), \\
t_m = \tanh \left( \frac{h-2 J_{\rm A} m+2 J_{\rm F} q}{2 T} \right).
\end{aligned}
\end{equation}

In the presence of the external magnetic field ($h \neq 0$), the necessary conditions for the energy minimum can be fulfilled for two solutions: (1) $(m_2+m_1)/2 = m$, describing an antiferromagnetic state (AFM, see Fig.\,\ref{fig:model}, bottom left panel),  and (2) $(m_2-m_1)/2 = q$, corresponding to field-polarized paramagnetism (PM, see Fig.\,\ref{fig:model}, bottom right panel) with the magnetization oriented along the field. By numerically solving the system of equations (\ref{eq:set-eqns}), we find the phase diagram, and field and temperature dependence of the average magnetization $M$ of four layers (\textit{i.e.} $M = \sum^4_{n=1} m_n/4$). In the numerical calculation we assume that $J_{\rm A}=1$, and the other interaction parameters, temperature and field are dimensionless, in units of $J_{\rm A}$.

Fig.~\ref{fig:theory}(a) shows the phase diagram $(h,T)$ with three regions: (i) AFM, where the antiferromagnetic solution satisfies the minimum conditions, (ii) META, where antiferromagnetic solution can exist as metastable, and (iii) PM, with the magnetization along the magnetic field. The field 
\begin{equation}
h = h_l = j_z - j_2, 
\end{equation} 
marks the boundary of the existence of the AFM phase, while the field 
\begin{equation}
h=h_m =-1/2( j_1+2 j_2)
\end{equation}
covers the region where the AFM solution yields a minimum. Below a certain value of the field, the metastable region almost disappears and probably cannot be observed experimentally. However, in this simplest molecular field approximation, the metastable state vanishes (and the system undergoes a continuous phase transition) only for $h=0$. Figs.~\ref{fig:theory}(b) and ~\ref{fig:theory}(c) show the field and temperature dependence of the magnetization, respectively, generated from the solution found. For small values of the field ($h<\frac{2}{9}$), the hysteresis is essentially invisible, which agrees with the conclusion regarding the vanishing of the metastable region.

Due to the assumptions of the model, we can compare the theoretical curves only with the experimental data collected for $H \parallel b$-axis, \textit{i.e.} the phase diagram in Fig.~\ref{fig:phase-diag}(b) to Fig.~\ref{fig:theory}(a), the field dependence of the magnetization in Figs.~\ref{fig:MvsH-b-c}(a,b) to Fig.~\ref{fig:theory}(b), and the temperature dependence of the magnetization in Fig.~\ref{fig:MvsT-all}(b) to Fig.~\ref{fig:theory}(c). Surprisingly, it turns out that the presented extremely simple model based on the molecular field approximation reproduces reasonably well some of the experimental observations. In particular, these are: (i) the presence of the antiferromagnetic and metastable state, (ii) vanishing magnetic hysteresis for temperature near $T_{\rm N}$ and low magnetic field, and (iii) a continuous phase transition in zero external magnetic field. Unfortunately, this model is not suitable for explaining any of the thermal expansion behavior.


\section{Summary and conclusions}

Summarizing, the properties of the single-crystalline compound UNi$_{0.34}$Ge$_2$ described in this work are in good agreement with the preliminary results reported for the polycrystalline sample of UNi$_{0.45}$Ge$_2$ \cite{Pasturel2021}. The observed small differences in the phase transition temperatures and fields of the polycrystalline and single-crystalline samples should be attributed solely to the impossible-to-avoid difference in their stoichiometry, due to the evidenced homogeneity domain for the UNi$_{1–x}$Ge$_2$ phase with $0.35 \leqslant x \leqslant 0.6$ \cite{Perricone2002}. The high quality of the crystal used in this study allowed for a distinct improvement in the sharpness of the antiferromagnetic and metamagnetic phase transitions. As a result, a reliable magnetic phase diagram was proposed for all three main crystallographic axes, constructed based on temperature- and field-dependent measurements of various thermodynamic properties.

Two observations seem most important in this study. First, given the magnetic structure of the system studied in zero magnetic field (\textit{i.e.} all the moments parallel or antiparallel to the $b$-axis) \cite{Pasturel2021}, one would naively expect UNi$_{0.34}$Ge$_2$ to have one easy magnetization axis (the $b$-axis), and that the $c$ axis to be as hard as the $a$ axis. Surprisingly, it turned out that UNi$_{0.34}$Ge$_2$ has two easy axes, $b$ and $c$, although the $b$-axis is indeed the easiest one. Only at the lowest temperatures and at a relatively low magnetic field do the $a$ and $c$ axes show some similarity in the magnetic behavior, well seen in the field dependence of magnetization (Fig.~\ref{fig:MvsH-b-c}). In other words, the magnetocrystalline anisotropy in UNi$_{0.34}$Ge$_2$ appears to be an intermediate case between the strong, uniaxial anisotropy (in which there is only one easy axis and two hard axes) and weak, cubic anisotropy (in which all axes show qualitatively the same behavior in the magnetic field). This clearly suggests the existence of some relationship between the observed anisotropy of the studied compound and the presence in its structure of uranium zig-zag chains aligned in the $bc$-plane.”

Second, we have shown that the magnetic structure of UNi$_{0.34}$Ge$_2$ in zero field and its temperature and field characteristics can be well described using a relatively simple, four-layer Ising model. Since this model considers only one, $z$-th component of the magnetic moments, which is parallel to the applied magnetic field direction, it obviously cannot explain the presence of the two easy axes in the system. Nevertheless, that simple approach shows that in zero field the layered structure of the system appears to be sufficient to lead to an antiferromagnetic collinear structure and the magnetic behavior observed experimentally.

Considering these observations, one can conclude that the zig-zag chains in UNi$_{0.34}$Ge$_2$ generally do not play a key role in the occurrence of magnetic ordering and its evolution in applied magnetic field. However, they seem to determine the anisotropy of the studied compound, especially the orientation of the uranium magnetic moments in $H\,\parallel\,c$, which presumably evolves towards that one observed in zero magnetic field in UGe$_2$, \textit{i.e.} the moments become parallel to the chains (\textit{cf.} Fig.\,\ref{fig:cryst-struct}). Further experiments (preferably neutron diffraction in magnetic fields) are needed to confirm this hypothesis. In turn, band structure calculations could shed more light on the possible role of Ni d-electron bands in the formation of the magnetic ground state of UNi$_{0.34}$Ge$_2$.


\begin{acknowledgments}
APP is grateful to the Kosciuszko Foundation -- the American Centre of Polish Culture (New York, NY, USA) for its financial support during his research stay at the Idaho National Laboratory. The work in Poland and France was co-financed by the Polish National Agency for Academic Exchange NAWA (Warsaw, Poland) and Campus France (Paris, France) within the PHC Polonium program (project no. PPN/BFR/2020/1/00022/U/00001). XD and KG acknowledge support from the INL's Laboratory Directed Research \& Development (LDRD) Program under DOE Idaho Operations Office Contract DE-AC07-05ID14517. The authors are grateful to V. Dorcet for his help in performing the TEM experiments on the THEMIS platform (ScanMAT, UAR 2025 University of Rennes\,1\,--\,CNRS; CPER-FEDER 2007–2014).
\end{acknowledgments}


\providecommand{\noopsort}[1]{}\providecommand{\singleletter}[1]{#1}%


\begin{thebibliography}{37}%
\makeatletter
\providecommand \@ifxundefined [1]{%
 \@ifx{#1\undefined}
}%
\providecommand \@ifnum [1]{%
 \ifnum #1\expandafter \@firstoftwo
 \else \expandafter \@secondoftwo
 \fi
}%
\providecommand \@ifx [1]{%
 \ifx #1\expandafter \@firstoftwo
 \else \expandafter \@secondoftwo
 \fi
}%
\providecommand \natexlab [1]{#1}%
\providecommand \enquote  [1]{``#1''}%
\providecommand \bibnamefont  [1]{#1}%
\providecommand \bibfnamefont [1]{#1}%
\providecommand \citenamefont [1]{#1}%
\providecommand \href@noop [0]{\@secondoftwo}%
\providecommand \href [0]{\begingroup \@sanitize@url \@href}%
\providecommand \@href[1]{\@@startlink{#1}\@@href}%
\providecommand \@@href[1]{\endgroup#1\@@endlink}%
\providecommand \@sanitize@url [0]{\catcode `\\12\catcode `\$12\catcode
  `\&12\catcode `\#12\catcode `\^12\catcode `\_12\catcode `\%12\relax}%
\providecommand \@@startlink[1]{}%
\providecommand \@@endlink[0]{}%
\providecommand \url  [0]{\begingroup\@sanitize@url \@url }%
\providecommand \@url [1]{\endgroup\@href {#1}{\urlprefix }}%
\providecommand \urlprefix  [0]{URL }%
\providecommand \Eprint [0]{\href }%
\providecommand \doibase [0]{https://doi.org/}%
\providecommand \selectlanguage [0]{\@gobble}%
\providecommand \bibinfo  [0]{\@secondoftwo}%
\providecommand \bibfield  [0]{\@secondoftwo}%
\providecommand \translation [1]{[#1]}%
\providecommand \BibitemOpen [0]{}%
\providecommand \bibitemStop [0]{}%
\providecommand \bibitemNoStop [0]{.\EOS\space}%
\providecommand \EOS [0]{\spacefactor3000\relax}%
\providecommand \BibitemShut  [1]{\csname bibitem#1\endcsname}%
\let\auto@bib@innerbib\@empty
\bibitem [{\citenamefont {Kim}(2012)}]{Kim2012}%
  \BibitemOpen
  \bibfield  {author} {\bibinfo {author} {\bibfnamefont {Y.~S.}\ \bibnamefont
  {Kim}},\ }\bibfield  {title} {\bibinfo {title} {3.14 - {U}ranium
  {I}ntermetallic {F}uels ({U}–{A}l, {U}–{S}i, {U}–{M}o)},\ }in\ \href
  {https://doi.org/doi.org/10.1016/B978-0-08-056033-5.00112-9} {\emph {\bibinfo
  {booktitle} {Comprehensive Nuclear Materials}}},\ \bibinfo {editor} {edited
  by\ \bibinfo {editor} {\bibfnamefont {R.~J.~M.}\ \bibnamefont {Konings}}}\
  (\bibinfo  {publisher} {Elsevier},\ \bibinfo {address} {Oxford},\ \bibinfo
  {year} {2012})\ pp.\ \bibinfo {pages} {391--422}\BibitemShut {NoStop}%
\bibitem [{\citenamefont {Ortega}\ \emph {et~al.}(2016)\citenamefont {Ortega},
  \citenamefont {Blamer}, \citenamefont {Evans},\ and\ \citenamefont
  {McDeavitt}}]{Ortega2016}%
  \BibitemOpen
  \bibfield  {author} {\bibinfo {author} {\bibfnamefont {L.~H.}\ \bibnamefont
  {Ortega}}, \bibinfo {author} {\bibfnamefont {B.~J.}\ \bibnamefont {Blamer}},
  \bibinfo {author} {\bibfnamefont {J.~A.}\ \bibnamefont {Evans}},\ and\
  \bibinfo {author} {\bibfnamefont {S.~M.}\ \bibnamefont {McDeavitt}},\
  }\bibfield  {title} {\bibinfo {title} {Development of an accident-tolerant
  fuel composite from uranium mononitride {(UN)} and uranium sesquisilicide
  {(U$_3$Si$_2$)} with increased uranium loading},\ }\href
  {https://doi.org/https://doi.org/10.1016/j.jnucmat.2016.01.014} {\bibfield
  {journal} {\bibinfo  {journal} {J. Nucl. Mater.}\ }\textbf {\bibinfo {volume}
  {471}},\ \bibinfo {pages} {116} (\bibinfo {year} {2016})}\BibitemShut
  {NoStop}%
\bibitem [{\citenamefont {Pfleiderer}(2009)}]{Pfleiderer2009}%
  \BibitemOpen
  \bibfield  {author} {\bibinfo {author} {\bibfnamefont {C.}~\bibnamefont
  {Pfleiderer}},\ }\bibfield  {title} {\bibinfo {title} {Superconducting phases
  of $f$-electron compounds},\ }\href
  {https://doi.org/10.1103/RevModPhys.81.1551} {\bibfield  {journal} {\bibinfo
  {journal} {Rev. Mod. Phys.}\ }\textbf {\bibinfo {volume} {81}},\ \bibinfo
  {pages} {1551} (\bibinfo {year} {2009})}\BibitemShut {NoStop}%
\bibitem [{\citenamefont {Ran}\ \emph {et~al.}(2019)\citenamefont {Ran},
  \citenamefont {Eckberg}, \citenamefont {Ding}, \citenamefont {Furukawa},
  \citenamefont {Metz}, \citenamefont {Saha}, \citenamefont {Liu},
  \citenamefont {Zic}, \citenamefont {Kim}, \citenamefont {Paglione},\ and\
  \citenamefont {Butch}}]{Ran2019}%
  \BibitemOpen
  \bibfield  {author} {\bibinfo {author} {\bibfnamefont {S.}~\bibnamefont
  {Ran}}, \bibinfo {author} {\bibfnamefont {C.}~\bibnamefont {Eckberg}},
  \bibinfo {author} {\bibfnamefont {Q.-P.}\ \bibnamefont {Ding}}, \bibinfo
  {author} {\bibfnamefont {Y.}~\bibnamefont {Furukawa}}, \bibinfo {author}
  {\bibfnamefont {T.}~\bibnamefont {Metz}}, \bibinfo {author} {\bibfnamefont
  {S.~R.}\ \bibnamefont {Saha}}, \bibinfo {author} {\bibfnamefont {I.-L.}\
  \bibnamefont {Liu}}, \bibinfo {author} {\bibfnamefont {M.}~\bibnamefont
  {Zic}}, \bibinfo {author} {\bibfnamefont {H.}~\bibnamefont {Kim}}, \bibinfo
  {author} {\bibfnamefont {J.}~\bibnamefont {Paglione}},\ and\ \bibinfo
  {author} {\bibfnamefont {N.~P.}\ \bibnamefont {Butch}},\ }\bibfield  {title}
  {\bibinfo {title} {Nearly ferromagnetic spin-triplet superconductivity},\
  }\href {https://doi.org/10.1126/science.aav8645} {\bibfield  {journal}
  {\bibinfo  {journal} {Science}\ }\textbf {\bibinfo {volume} {365}},\ \bibinfo
  {pages} {684} (\bibinfo {year} {2019})}\BibitemShut {NoStop}%
\bibitem [{\citenamefont {Aoki}\ \emph
  {et~al.}(2019{\natexlab{a}})\citenamefont {Aoki}, \citenamefont {Nakamura},
  \citenamefont {Honda}, \citenamefont {Li}, \citenamefont {Homma},
  \citenamefont {Shimizu}, \citenamefont {Sato}, \citenamefont {Knebel},
  \citenamefont {Brison}, \citenamefont {Pourret}, \citenamefont {Braithwaite},
  \citenamefont {Lapertot}, \citenamefont {Niu}, \citenamefont {Vališka},
  \citenamefont {Harima},\ and\ \citenamefont {Flouquet}}]{Aoki2019a}%
  \BibitemOpen
  \bibfield  {author} {\bibinfo {author} {\bibfnamefont {D.}~\bibnamefont
  {Aoki}}, \bibinfo {author} {\bibfnamefont {A.}~\bibnamefont {Nakamura}},
  \bibinfo {author} {\bibfnamefont {F.}~\bibnamefont {Honda}}, \bibinfo
  {author} {\bibfnamefont {D.}~\bibnamefont {Li}}, \bibinfo {author}
  {\bibfnamefont {Y.}~\bibnamefont {Homma}}, \bibinfo {author} {\bibfnamefont
  {Y.}~\bibnamefont {Shimizu}}, \bibinfo {author} {\bibfnamefont {Y.~J.}\
  \bibnamefont {Sato}}, \bibinfo {author} {\bibfnamefont {G.}~\bibnamefont
  {Knebel}}, \bibinfo {author} {\bibfnamefont {J.-P.}\ \bibnamefont {Brison}},
  \bibinfo {author} {\bibfnamefont {A.}~\bibnamefont {Pourret}}, \bibinfo
  {author} {\bibfnamefont {D.}~\bibnamefont {Braithwaite}}, \bibinfo {author}
  {\bibfnamefont {G.}~\bibnamefont {Lapertot}}, \bibinfo {author}
  {\bibfnamefont {Q.}~\bibnamefont {Niu}}, \bibinfo {author} {\bibfnamefont
  {M.}~\bibnamefont {Vališka}}, \bibinfo {author} {\bibfnamefont
  {H.}~\bibnamefont {Harima}},\ and\ \bibinfo {author} {\bibfnamefont
  {J.}~\bibnamefont {Flouquet}},\ }\bibfield  {title} {\bibinfo {title}
  {Unconventional {S}uperconductivity in {H}eavy {F}ermion {UT}e$_2$},\ }\href
  {https://doi.org/10.7566/JPSJ.88.043702} {\bibfield  {journal} {\bibinfo
  {journal} {J. Phys. Soc. Japan}\ }\textbf {\bibinfo {volume} {88}},\ \bibinfo
  {pages} {043702} (\bibinfo {year} {2019}{\natexlab{a}})}\BibitemShut
  {NoStop}%
\bibitem [{\citenamefont {Kaczorowski}\ and\ \citenamefont
  {Tro{\'c}}(1990)}]{Kaczorowski1990}%
  \BibitemOpen
  \bibfield  {author} {\bibinfo {author} {\bibfnamefont {D.}~\bibnamefont
  {Kaczorowski}}\ and\ \bibinfo {author} {\bibfnamefont {R.}~\bibnamefont
  {Tro{\'c}}},\ }\bibfield  {title} {\bibinfo {title} {Magnetic and transport
  properties of a strongly anisotropic ferromagnet, {UCu$_2$P$_2$}},\ }\href
  {https://doi.org/doi.org/10.1088/0953-8984/2/18/015} {\bibfield  {journal}
  {\bibinfo  {journal} {J. Phys.: Condens. Matter}\ }\textbf {\bibinfo {volume}
  {2}},\ \bibinfo {pages} {4185} (\bibinfo {year} {1990})}\BibitemShut
  {NoStop}%
\bibitem [{\citenamefont {Stali{\'n}ski}(1974)}]{Stalinski1974}%
  \BibitemOpen
  \bibfield  {author} {\bibinfo {author} {\bibfnamefont {B.}~\bibnamefont
  {Stali{\'n}ski}},\ }\bibfield  {title} {\bibinfo {title} {Magnetic properties
  of some simple metallic and semimetallic compounds of f‐electron metals
  with main group elements},\ }\href
  {https://doi.org/doi.org/10.1063/1.3141760} {\bibfield  {journal} {\bibinfo
  {journal} {AIP Conf. Proc.}\ }\textbf {\bibinfo {volume} {18}},\ \bibinfo
  {pages} {490} (\bibinfo {year} {1974})}\BibitemShut {NoStop}%
\bibitem [{\citenamefont {Zwicknagl}\ and\ \citenamefont
  {Fulde}(2003)}]{Zwicknagl2003}%
  \BibitemOpen
  \bibfield  {author} {\bibinfo {author} {\bibfnamefont {G.}~\bibnamefont
  {Zwicknagl}}\ and\ \bibinfo {author} {\bibfnamefont {P.}~\bibnamefont
  {Fulde}},\ }\bibfield  {title} {\bibinfo {title} {The dual nature of 5f
  electrons and the origin of heavy fermions in {U} compounds},\ }\href
  {https://doi.org/doi.org/10.1088/0953-8984/15/28/302} {\bibfield  {journal}
  {\bibinfo  {journal} {J. of Phys.: Condens. Matter}\ }\textbf {\bibinfo
  {volume} {15}},\ \bibinfo {pages} {S1911} (\bibinfo {year}
  {2003})}\BibitemShut {NoStop}%
\bibitem [{\citenamefont {Tro\'{c}}\ \emph {et~al.}(2012)\citenamefont
  {Tro\'{c}}, \citenamefont {Gajek},\ and\ \citenamefont {Pikul}}]{Troc2012}%
  \BibitemOpen
  \bibfield  {author} {\bibinfo {author} {\bibfnamefont {R.}~\bibnamefont
  {Tro\'{c}}}, \bibinfo {author} {\bibfnamefont {Z.}~\bibnamefont {Gajek}},\
  and\ \bibinfo {author} {\bibfnamefont {A.}~\bibnamefont {Pikul}},\ }\bibfield
   {title} {\bibinfo {title} {Dualism of the 5f electrons of the ferromagnetic
  superconductor {UGe$_{2}$} as seen in magnetic, transport, and specific-heat
  data},\ }\href {https://doi.org/10.1103/PhysRevB.86.224403} {\bibfield
  {journal} {\bibinfo  {journal} {Phys. Rev. B}\ }\textbf {\bibinfo {volume}
  {86}},\ \bibinfo {pages} {224403} (\bibinfo {year} {2012})}\BibitemShut
  {NoStop}%
\bibitem [{\citenamefont {Lee}\ \emph {et~al.}(2018)\citenamefont {Lee},
  \citenamefont {Matsuda}, \citenamefont {Mydosh}, \citenamefont {Zaliznyak},
  \citenamefont {Kolesnikov}, \citenamefont {S\"ullow}, \citenamefont {Ruff},\
  and\ \citenamefont {Granroth}}]{Lee2018}%
  \BibitemOpen
  \bibfield  {author} {\bibinfo {author} {\bibfnamefont {J.}~\bibnamefont
  {Lee}}, \bibinfo {author} {\bibfnamefont {M.}~\bibnamefont {Matsuda}},
  \bibinfo {author} {\bibfnamefont {J.~A.}\ \bibnamefont {Mydosh}}, \bibinfo
  {author} {\bibfnamefont {I.}~\bibnamefont {Zaliznyak}}, \bibinfo {author}
  {\bibfnamefont {A.~I.}\ \bibnamefont {Kolesnikov}}, \bibinfo {author}
  {\bibfnamefont {S.}~\bibnamefont {S\"ullow}}, \bibinfo {author}
  {\bibfnamefont {J.~P.~C.}\ \bibnamefont {Ruff}},\ and\ \bibinfo {author}
  {\bibfnamefont {G.~E.}\ \bibnamefont {Granroth}},\ }\bibfield  {title}
  {\bibinfo {title} {Dual {N}ature of {M}agnetism in a {U}ranium
  {H}eavy-{F}ermion {S}ystem},\ }\href
  {https://doi.org/10.1103/PhysRevLett.121.057201} {\bibfield  {journal}
  {\bibinfo  {journal} {Phys. Rev. Lett.}\ }\textbf {\bibinfo {volume} {121}},\
  \bibinfo {pages} {057201} (\bibinfo {year} {2018})}\BibitemShut {NoStop}%
\bibitem [{\citenamefont {Amorese}\ \emph {et~al.}(2020)\citenamefont
  {Amorese}, \citenamefont {Sundermann}, \citenamefont {Leedahl}, \citenamefont
  {Marino}, \citenamefont {Takegami}, \citenamefont {Gretarsson}, \citenamefont
  {Gloskovskii}, \citenamefont {Schlueter}, \citenamefont {Haverkort},
  \citenamefont {Huang}, \citenamefont {Szlawska}, \citenamefont {Kaczorowski},
  \citenamefont {Ran}, \citenamefont {Maple}, \citenamefont {Bauer},
  \citenamefont {Leithe-Jasper}, \citenamefont {Hansmann}, \citenamefont
  {Thalmeier}, \citenamefont {Tjeng},\ and\ \citenamefont
  {Severing}}]{Amorese2020}%
  \BibitemOpen
  \bibfield  {author} {\bibinfo {author} {\bibfnamefont {A.}~\bibnamefont
  {Amorese}}, \bibinfo {author} {\bibfnamefont {M.}~\bibnamefont {Sundermann}},
  \bibinfo {author} {\bibfnamefont {B.}~\bibnamefont {Leedahl}}, \bibinfo
  {author} {\bibfnamefont {A.}~\bibnamefont {Marino}}, \bibinfo {author}
  {\bibfnamefont {D.}~\bibnamefont {Takegami}}, \bibinfo {author}
  {\bibfnamefont {H.}~\bibnamefont {Gretarsson}}, \bibinfo {author}
  {\bibfnamefont {A.}~\bibnamefont {Gloskovskii}}, \bibinfo {author}
  {\bibfnamefont {C.}~\bibnamefont {Schlueter}}, \bibinfo {author}
  {\bibfnamefont {M.~W.}\ \bibnamefont {Haverkort}}, \bibinfo {author}
  {\bibfnamefont {Y.}~\bibnamefont {Huang}}, \bibinfo {author} {\bibfnamefont
  {M.}~\bibnamefont {Szlawska}}, \bibinfo {author} {\bibfnamefont
  {D.}~\bibnamefont {Kaczorowski}}, \bibinfo {author} {\bibfnamefont
  {S.}~\bibnamefont {Ran}}, \bibinfo {author} {\bibfnamefont {M.~B.}\
  \bibnamefont {Maple}}, \bibinfo {author} {\bibfnamefont {E.~D.}\ \bibnamefont
  {Bauer}}, \bibinfo {author} {\bibfnamefont {A.}~\bibnamefont
  {Leithe-Jasper}}, \bibinfo {author} {\bibfnamefont {P.}~\bibnamefont
  {Hansmann}}, \bibinfo {author} {\bibfnamefont {P.}~\bibnamefont {Thalmeier}},
  \bibinfo {author} {\bibfnamefont {L.~H.}\ \bibnamefont {Tjeng}},\ and\
  \bibinfo {author} {\bibfnamefont {A.}~\bibnamefont {Severing}},\ }\bibfield
  {title} {\bibinfo {title} {From antiferromagnetic and hidden order to {P}auli
  paramagnetism in {U}{$M_2$}{S}i$_2$ compounds with 5f electron duality},\
  }\href {https://doi.org/10.1073/pnas.2005701117} {\bibfield  {journal}
  {\bibinfo  {journal} {Proc. National Academy of Sciences}\ }\textbf {\bibinfo
  {volume} {117}},\ \bibinfo {pages} {30220} (\bibinfo {year}
  {2020})}\BibitemShut {NoStop}%
\bibitem [{\citenamefont {Saxena}\ \emph {et~al.}(2000)\citenamefont {Saxena},
  \citenamefont {Agarwal}, \citenamefont {Ahilan}, \citenamefont {Grosche},
  \citenamefont {Haselwimmer}, \citenamefont {Steiner}, \citenamefont {Pugh},
  \citenamefont {Walker}, \citenamefont {Julian}, \citenamefont {Monthoux},
  \citenamefont {Lonzarich}, \citenamefont {Huxley}, \citenamefont {Sheikin},
  \citenamefont {Braithwaite},\ and\ \citenamefont {Flouquet}}]{Saxena2000}%
  \BibitemOpen
  \bibfield  {author} {\bibinfo {author} {\bibfnamefont {S.~S.}\ \bibnamefont
  {Saxena}}, \bibinfo {author} {\bibfnamefont {P.}~\bibnamefont {Agarwal}},
  \bibinfo {author} {\bibfnamefont {K.}~\bibnamefont {Ahilan}}, \bibinfo
  {author} {\bibfnamefont {F.~M.}\ \bibnamefont {Grosche}}, \bibinfo {author}
  {\bibfnamefont {R.~K.~W.}\ \bibnamefont {Haselwimmer}}, \bibinfo {author}
  {\bibfnamefont {M.~J.}\ \bibnamefont {Steiner}}, \bibinfo {author}
  {\bibfnamefont {E.}~\bibnamefont {Pugh}}, \bibinfo {author} {\bibfnamefont
  {I.~R.}\ \bibnamefont {Walker}}, \bibinfo {author} {\bibfnamefont {S.~R.}\
  \bibnamefont {Julian}}, \bibinfo {author} {\bibfnamefont {P.}~\bibnamefont
  {Monthoux}}, \bibinfo {author} {\bibfnamefont {G.~G.}\ \bibnamefont
  {Lonzarich}}, \bibinfo {author} {\bibfnamefont {A.}~\bibnamefont {Huxley}},
  \bibinfo {author} {\bibfnamefont {I.}~\bibnamefont {Sheikin}}, \bibinfo
  {author} {\bibfnamefont {D.}~\bibnamefont {Braithwaite}},\ and\ \bibinfo
  {author} {\bibfnamefont {J.}~\bibnamefont {Flouquet}},\ }\bibfield  {title}
  {\bibinfo {title} {Superconductivity on the border of itinerant-electron
  ferromagnetism in {U}{Ge}$_2$},\ }\href {https://doi.org/10.1038/35020500}
  {\bibfield  {journal} {\bibinfo  {journal} {Nature}\ }\textbf {\bibinfo
  {volume} {406}},\ \bibinfo {pages} {587–592} (\bibinfo {year}
  {2000})}\BibitemShut {NoStop}%
\bibitem [{\citenamefont {Aoki}\ \emph {et~al.}(2001)\citenamefont {Aoki},
  \citenamefont {Huxley}, \citenamefont {Ressouche}, \citenamefont
  {Braithwaite}, \citenamefont {Flouquet}, \citenamefont {Brison},
  \citenamefont {Lhotel},\ and\ \citenamefont {Paulsen}}]{Aoki2001}%
  \BibitemOpen
  \bibfield  {author} {\bibinfo {author} {\bibfnamefont {D.}~\bibnamefont
  {Aoki}}, \bibinfo {author} {\bibfnamefont {A.}~\bibnamefont {Huxley}},
  \bibinfo {author} {\bibfnamefont {E.}~\bibnamefont {Ressouche}}, \bibinfo
  {author} {\bibfnamefont {D.}~\bibnamefont {Braithwaite}}, \bibinfo {author}
  {\bibfnamefont {J.}~\bibnamefont {Flouquet}}, \bibinfo {author}
  {\bibfnamefont {J.-P.}\ \bibnamefont {Brison}}, \bibinfo {author}
  {\bibfnamefont {E.}~\bibnamefont {Lhotel}},\ and\ \bibinfo {author}
  {\bibfnamefont {C.}~\bibnamefont {Paulsen}},\ }\bibfield  {title} {\bibinfo
  {title} {Coexistence of superconductivity and ferromagnetism in
  {U}{R}h{G}e},\ }\href {https://doi.org/10.1038/35098048} {\bibfield
  {journal} {\bibinfo  {journal} {Nature}\ }\textbf {\bibinfo {volume} {413}},\
  \bibinfo {pages} {613 – 616} (\bibinfo {year} {2001})}\BibitemShut
  {NoStop}%
\bibitem [{\citenamefont {Huy}\ \emph {et~al.}(2007)\citenamefont {Huy},
  \citenamefont {Gasparini}, \citenamefont {de~Nijs}, \citenamefont {Huang},
  \citenamefont {Klaasse}, \citenamefont {Gortenmulder}, \citenamefont
  {de~Visser}, \citenamefont {Hamann}, \citenamefont {G\"orlach},\ and\
  \citenamefont {L\"ohneysen}}]{Huy2007}%
  \BibitemOpen
  \bibfield  {author} {\bibinfo {author} {\bibfnamefont {N.~T.}\ \bibnamefont
  {Huy}}, \bibinfo {author} {\bibfnamefont {A.}~\bibnamefont {Gasparini}},
  \bibinfo {author} {\bibfnamefont {D.~E.}\ \bibnamefont {de~Nijs}}, \bibinfo
  {author} {\bibfnamefont {Y.}~\bibnamefont {Huang}}, \bibinfo {author}
  {\bibfnamefont {J.~C.~P.}\ \bibnamefont {Klaasse}}, \bibinfo {author}
  {\bibfnamefont {T.}~\bibnamefont {Gortenmulder}}, \bibinfo {author}
  {\bibfnamefont {A.}~\bibnamefont {de~Visser}}, \bibinfo {author}
  {\bibfnamefont {A.}~\bibnamefont {Hamann}}, \bibinfo {author} {\bibfnamefont
  {T.}~\bibnamefont {G\"orlach}},\ and\ \bibinfo {author} {\bibfnamefont
  {H.~v.}\ \bibnamefont {L\"ohneysen}},\ }\bibfield  {title} {\bibinfo {title}
  {Superconductivity on the {B}order of {W}eak {I}tinerant {F}erromagnetism in
  {U}{C}o{G}e},\ }\href {https://doi.org/10.1103/PhysRevLett.99.067006}
  {\bibfield  {journal} {\bibinfo  {journal} {Phys. Rev. Lett.}\ }\textbf
  {\bibinfo {volume} {99}},\ \bibinfo {pages} {067006} (\bibinfo {year}
  {2007})}\BibitemShut {NoStop}%
\bibitem [{\citenamefont {Henriques}\ \emph {et~al.}(2015)\citenamefont
  {Henriques}, \citenamefont {Berthebaud}, \citenamefont {Lignie},
  \citenamefont {{El Sayah}}, \citenamefont {Moussa}, \citenamefont {Tougait},
  \citenamefont {Havela},\ and\ \citenamefont
  {Gon{\c{c}}alves}}]{Henriques2015}%
  \BibitemOpen
  \bibfield  {author} {\bibinfo {author} {\bibfnamefont {M.~S.}\ \bibnamefont
  {Henriques}}, \bibinfo {author} {\bibfnamefont {D.}~\bibnamefont
  {Berthebaud}}, \bibinfo {author} {\bibfnamefont {A.}~\bibnamefont {Lignie}},
  \bibinfo {author} {\bibfnamefont {Z.}~\bibnamefont {{El Sayah}}}, \bibinfo
  {author} {\bibfnamefont {C.}~\bibnamefont {Moussa}}, \bibinfo {author}
  {\bibfnamefont {O.}~\bibnamefont {Tougait}}, \bibinfo {author} {\bibfnamefont
  {L.}~\bibnamefont {Havela}},\ and\ \bibinfo {author} {\bibfnamefont
  {A.}~\bibnamefont {Gon{\c{c}}alves}},\ }\bibfield  {title} {\bibinfo {title}
  {Isothermal section of the ternary phase diagram {U}–{Fe}–{Ge} at
  900$^{\circ}${C} and its new intermetallic phases},\ }\href
  {https://doi.org/https://doi.org/10.1016/j.jallcom.2015.03.145} {\bibfield
  {journal} {\bibinfo  {journal} {J. Alloys Compd.}\ }\textbf {\bibinfo
  {volume} {639}},\ \bibinfo {pages} {224} (\bibinfo {year}
  {2015})}\BibitemShut {NoStop}%
\bibitem [{\citenamefont {Szlawska}\ \emph {et~al.}(2022)\citenamefont
  {Szlawska}, \citenamefont {Pasturel}, \citenamefont {Kaczorowski},\ and\
  \citenamefont {Pikul}}]{Szlawska2022}%
  \BibitemOpen
  \bibfield  {author} {\bibinfo {author} {\bibfnamefont {M.}~\bibnamefont
  {Szlawska}}, \bibinfo {author} {\bibfnamefont {M.}~\bibnamefont {Pasturel}},
  \bibinfo {author} {\bibfnamefont {D.}~\bibnamefont {Kaczorowski}},\ and\
  \bibinfo {author} {\bibfnamefont {A.}~\bibnamefont {Pikul}},\ }\bibfield
  {title} {\bibinfo {title} {Ferromagnetism in structurally disordered
  {U}{Fe}$_{0.39}${Ge}$_2$},\ }\href
  {https://doi.org/https://doi.org/10.1016/j.jallcom.2021.162032} {\bibfield
  {journal} {\bibinfo  {journal} {J. Alloys Compd.}\ }\textbf {\bibinfo
  {volume} {892}},\ \bibinfo {pages} {162032} (\bibinfo {year}
  {2022})}\BibitemShut {NoStop}%
\bibitem [{\citenamefont {Pikul}\ \emph {et~al.}(2022)\citenamefont {Pikul},
  \citenamefont {Idczak}, \citenamefont {Sobota}, \citenamefont {Nowak},
  \citenamefont {Pasturel},\ and\ \citenamefont {Tran}}]{Pikul2022}%
  \BibitemOpen
  \bibfield  {author} {\bibinfo {author} {\bibfnamefont {A.~P.}\ \bibnamefont
  {Pikul}}, \bibinfo {author} {\bibfnamefont {R.}~\bibnamefont {Idczak}},
  \bibinfo {author} {\bibfnamefont {P.}~\bibnamefont {Sobota}}, \bibinfo
  {author} {\bibfnamefont {W.}~\bibnamefont {Nowak}}, \bibinfo {author}
  {\bibfnamefont {M.}~\bibnamefont {Pasturel}},\ and\ \bibinfo {author}
  {\bibfnamefont {V.~H.}\ \bibnamefont {Tran}},\ }\bibfield  {title} {\bibinfo
  {title} {Ferromagnetic ordering in {U}{Fe}$_{0.40}${Ge}$_2$ studied by
  $^{57}${Fe} {M}össbauer spectroscopy},\ }\href
  {https://doi.org/https://doi.org/10.1016/j.jmmm.2022.169238} {\bibfield
  {journal} {\bibinfo  {journal} {J. Magn. Magn. Mater.}\ }\textbf {\bibinfo
  {volume} {553}},\ \bibinfo {pages} {169238} (\bibinfo {year}
  {2022})}\BibitemShut {NoStop}%
\bibitem [{\citenamefont {Soud{\'e}}(2010)}]{Soude2010}%
  \BibitemOpen
  \bibfield  {author} {\bibinfo {author} {\bibfnamefont {A.}~\bibnamefont
  {Soud{\'e}}},\ }\href@noop {} {} (\bibinfo {year} {2010}),\ \bibinfo {note}
  {{C}aractérisations chimiques, structurales et électroniques des phases
  intermétalliques ({Ce},{U})-{Co}-{Ge}, Doctoral dissertation (in {F}rench),
  {U}niversit{\'e} {R}ennes 1, Renens, France}\BibitemShut {NoStop}%
\bibitem [{\citenamefont {Molcanova}\ \emph {et~al.}(2017)\citenamefont
  {Molcanova}, \citenamefont {Mihalik}, \citenamefont {andM. Reiffers},
  \citenamefont {Dzubinska}, \citenamefont {Hurakova}, \citenamefont
  {Kavecansky}, \citenamefont {Paukov},\ and\ \citenamefont
  {Havela}}]{Molcanova2017}%
  \BibitemOpen
  \bibfield  {author} {\bibinfo {author} {\bibfnamefont {Z.}~\bibnamefont
  {Molcanova}}, \bibinfo {author} {\bibfnamefont {M.}~\bibnamefont {Mihalik}},
  \bibinfo {author} {\bibfnamefont {M.~M.~J.}\ \bibnamefont {andM. Reiffers}},
  \bibinfo {author} {\bibfnamefont {A.}~\bibnamefont {Dzubinska}}, \bibinfo
  {author} {\bibfnamefont {M.}~\bibnamefont {Hurakova}}, \bibinfo {author}
  {\bibfnamefont {V.}~\bibnamefont {Kavecansky}}, \bibinfo {author}
  {\bibfnamefont {M.}~\bibnamefont {Paukov}},\ and\ \bibinfo {author}
  {\bibfnamefont {L.}~\bibnamefont {Havela}},\ }\bibfield  {title} {\bibinfo
  {title} {Characterization of {N}ew {U}-{Ni}-{X}$_2$ {S}plats and {S}tudy of
  their {P}hysical {P}roperties},\ }\href
  {https://doi.org/https://doi.org/10.12693/APhysPolA.131.994} {\bibfield
  {journal} {\bibinfo  {journal} {Acta Phys. Pol. A}\ }\textbf {\bibinfo
  {volume} {131}},\ \bibinfo {pages} {994} (\bibinfo {year}
  {2017})}\BibitemShut {NoStop}%
\bibitem [{\citenamefont {Ohashi}\ \emph {et~al.}(2018)\citenamefont {Ohashi},
  \citenamefont {Ohashi}, \citenamefont {Sawabu}, \citenamefont {Miyagawa},
  \citenamefont {Maeta},\ and\ \citenamefont {Yamamura}}]{Ohashi2018}%
  \BibitemOpen
  \bibfield  {author} {\bibinfo {author} {\bibfnamefont {K.}~\bibnamefont
  {Ohashi}}, \bibinfo {author} {\bibfnamefont {M.}~\bibnamefont {Ohashi}},
  \bibinfo {author} {\bibfnamefont {M.}~\bibnamefont {Sawabu}}, \bibinfo
  {author} {\bibfnamefont {M.}~\bibnamefont {Miyagawa}}, \bibinfo {author}
  {\bibfnamefont {K.}~\bibnamefont {Maeta}},\ and\ \bibinfo {author}
  {\bibfnamefont {T.}~\bibnamefont {Yamamura}},\ }\bibfield  {title} {\bibinfo
  {title} {Magnetic properties of the {U}{Ni}{Ge}$_2$ at low temperature},\
  }\href {https://doi.org/https://doi.org/10.1088/1742-6596/969/1/012101}
  {\bibfield  {journal} {\bibinfo  {journal} {J. Phys.: Conf. Ser.}\ }\textbf
  {\bibinfo {volume} {969}},\ \bibinfo {pages} {012101} (\bibinfo {year}
  {2018})}\BibitemShut {NoStop}%
\bibitem [{\citenamefont {Pasturel}\ \emph {et~al.}(2021)\citenamefont
  {Pasturel}, \citenamefont {Szlawska}, \citenamefont {{\'C}wik}, \citenamefont
  {Kaczorowski},\ and\ \citenamefont {Pikul}}]{Pasturel2021}%
  \BibitemOpen
  \bibfield  {author} {\bibinfo {author} {\bibfnamefont {M.}~\bibnamefont
  {Pasturel}}, \bibinfo {author} {\bibfnamefont {M.}~\bibnamefont {Szlawska}},
  \bibinfo {author} {\bibfnamefont {J.}~\bibnamefont {{\'C}wik}}, \bibinfo
  {author} {\bibfnamefont {D.}~\bibnamefont {Kaczorowski}},\ and\ \bibinfo
  {author} {\bibfnamefont {A.~P.}\ \bibnamefont {Pikul}},\ }\bibfield  {title}
  {\bibinfo {title} {Antiferromagnetic ordering in the ternary uranium
  germanide {U}{Ni}$_{1-x}${Ge}$_2$: Neutron diffraction and physical
  properties studies},\ }\href
  {https://doi.org/https://doi.org/10.1016/j.intermet.2021.107112} {\bibfield
  {journal} {\bibinfo  {journal} {Intermetallics}\ }\textbf {\bibinfo {volume}
  {131}},\ \bibinfo {pages} {107112} (\bibinfo {year} {2021})}\BibitemShut
  {NoStop}%
\bibitem [{\citenamefont {Pasturel}\ \emph {et~al.}(2018)\citenamefont
  {Pasturel}, \citenamefont {Pikul}, \citenamefont {Chajewski}, \citenamefont
  {No{\"e}l},\ and\ \citenamefont {Kaczorowski}}]{Pasturel2018}%
  \BibitemOpen
  \bibfield  {author} {\bibinfo {author} {\bibfnamefont {M.}~\bibnamefont
  {Pasturel}}, \bibinfo {author} {\bibfnamefont {A.}~\bibnamefont {Pikul}},
  \bibinfo {author} {\bibfnamefont {G.}~\bibnamefont {Chajewski}}, \bibinfo
  {author} {\bibfnamefont {H.}~\bibnamefont {No{\"e}l}},\ and\ \bibinfo
  {author} {\bibfnamefont {D.}~\bibnamefont {Kaczorowski}},\ }\bibfield
  {title} {\bibinfo {title} {Ferromagnetic ordering in the novel ternary
  uranium germanide {U}{Ru}$_{0.29}${Ge}$_2$},\ }\href
  {https://doi.org/https://doi.org/10.1016/j.intermet.2018.01.011} {\bibfield
  {journal} {\bibinfo  {journal} {Intermetallics}\ }\textbf {\bibinfo {volume}
  {95}},\ \bibinfo {pages} {19} (\bibinfo {year} {2018})}\BibitemShut {NoStop}%
\bibitem [{\citenamefont {Pikul}(2019)}]{Pikul2019JdA}%
  \BibitemOpen
  \bibfield  {author} {\bibinfo {author} {\bibfnamefont {A.~P.}\ \bibnamefont
  {Pikul}},\ }\href@noop {} {} (\bibinfo {year} {2019}),\ \bibinfo {note}
  {magnetic and related properties of novel phases {U}{T}$_{1–x}${Ge}$_2$
  ({T} = {Fe}, {Ni}, {Os}), in: {A}bstracts of the 49{\'e}mes {J}ourn{\'e}es
  des {A}ctinides, {A}pril 14-18, 2019, {E}rice, {I}taly}\BibitemShut {NoStop}%
\bibitem [{\citenamefont {Aoki}\ \emph
  {et~al.}(2019{\natexlab{b}})\citenamefont {Aoki}, \citenamefont {Ishida},\
  and\ \citenamefont {Flouquet}}]{Aoki2019}%
  \BibitemOpen
  \bibfield  {author} {\bibinfo {author} {\bibfnamefont {D.}~\bibnamefont
  {Aoki}}, \bibinfo {author} {\bibfnamefont {K.}~\bibnamefont {Ishida}},\ and\
  \bibinfo {author} {\bibfnamefont {J.}~\bibnamefont {Flouquet}},\ }\bibfield
  {title} {\bibinfo {title} {Review of {U}-based {F}erromagnetic
  {S}uperconductors: {C}omparison between {U}{G}e$_2$, {U}{R}h{G}e, and
  {U}{C}o{G}e},\ }\href {https://doi.org/10.7566/JPSJ.88.022001} {\bibfield
  {journal} {\bibinfo  {journal} {J. Phys. Soc. Jpn.}\ }\textbf {\bibinfo
  {volume} {88}},\ \bibinfo {pages} {022001} (\bibinfo {year}
  {2019}{\natexlab{b}})}\BibitemShut {NoStop}%
\bibitem [{\citenamefont {Sheldrick}(2015)}]{Sheldrick2015}%
  \BibitemOpen
  \bibfield  {author} {\bibinfo {author} {\bibfnamefont {G.~M.}\ \bibnamefont
  {Sheldrick}},\ }\bibfield  {title} {\bibinfo {title} {{Crystal structure
  refinement with {SHELXL}}},\ }\href
  {https://doi.org/10.1107/S2053229614024218} {\bibfield  {journal} {\bibinfo
  {journal} {Acta Crystallogr., Sect. C: Cryst. Struct. Commun.}\ }\textbf
  {\bibinfo {volume} {71}},\ \bibinfo {pages} {3} (\bibinfo {year}
  {2015})}\BibitemShut {NoStop}%
\bibitem [{\citenamefont {Dolomanov}\ \emph {et~al.}(2009)\citenamefont
  {Dolomanov}, \citenamefont {Bourhis}, \citenamefont {Gildea}, \citenamefont
  {Howard},\ and\ \citenamefont {Puschmann}}]{Dolomanov2009}%
  \BibitemOpen
  \bibfield  {author} {\bibinfo {author} {\bibfnamefont {O.~V.}\ \bibnamefont
  {Dolomanov}}, \bibinfo {author} {\bibfnamefont {L.~J.}\ \bibnamefont
  {Bourhis}}, \bibinfo {author} {\bibfnamefont {R.~J.}\ \bibnamefont {Gildea}},
  \bibinfo {author} {\bibfnamefont {J.~A.~K.}\ \bibnamefont {Howard}},\ and\
  \bibinfo {author} {\bibfnamefont {H.}~\bibnamefont {Puschmann}},\ }\bibfield
  {title} {\bibinfo {title} {{{OLEX2}: a complete structure solution,
  refinement and analysis program}},\ }\href
  {https://doi.org/10.1107/S0021889808042726} {\bibfield  {journal} {\bibinfo
  {journal} {J. Appl. Crystallogr.}\ }\textbf {\bibinfo {volume} {42}},\
  \bibinfo {pages} {339} (\bibinfo {year} {2009})}\BibitemShut {NoStop}%
\bibitem [{\citenamefont {Zhuravleva}\ \emph {et~al.}(2005)\citenamefont
  {Zhuravleva}, \citenamefont {Bilc}, \citenamefont {Pcionek}, \citenamefont
  {Mahanti},\ and\ \citenamefont {Kanatzidis}}]{Zuravleva2005}%
  \BibitemOpen
  \bibfield  {author} {\bibinfo {author} {\bibfnamefont {M.~A.}\ \bibnamefont
  {Zhuravleva}}, \bibinfo {author} {\bibfnamefont {D.}~\bibnamefont {Bilc}},
  \bibinfo {author} {\bibfnamefont {R.~J.}\ \bibnamefont {Pcionek}}, \bibinfo
  {author} {\bibfnamefont {S.~D.}\ \bibnamefont {Mahanti}},\ and\ \bibinfo
  {author} {\bibfnamefont {M.~G.}\ \bibnamefont {Kanatzidis}},\ }\bibfield
  {title} {\bibinfo {title} {{Tb}$_4${Fe}{Ge}$_8$ grown in liquid {G}allium:
  {Trans-Cis} chains from the distortion of a planar {Ge} square net},\ }\href
  {https://doi.org/https://doi.org/10.1021/ic0487878} {\bibfield  {journal}
  {\bibinfo  {journal} {Inorg. Chem.}\ }\textbf {\bibinfo {volume} {44}},\
  \bibinfo {pages} {2177–2188} (\bibinfo {year} {2005})}\BibitemShut
  {NoStop}%
\bibitem [{\citenamefont {Zhang}\ \emph {et~al.}(2015)\citenamefont {Zhang},
  \citenamefont {Wang},\ and\ \citenamefont {Bobev}}]{Zhang2015}%
  \BibitemOpen
  \bibfield  {author} {\bibinfo {author} {\bibfnamefont {J.}~\bibnamefont
  {Zhang}}, \bibinfo {author} {\bibfnamefont {Y.}~\bibnamefont {Wang}},\ and\
  \bibinfo {author} {\bibfnamefont {S.}~\bibnamefont {Bobev}},\ }\bibfield
  {title} {\bibinfo {title} {Structural {M}odulations in the {R}are-{E}arth
  {M}etal {D}igermanides {REAl}$_{1–x}${Ge}$_2$ ({RE} = {Gd–Tm}, {Lu}, {Y};
  0.8 $<$ x $<$ 0.9). {C}orrelations between {L}ong- and {S}hort-{R}ange
  {V}acancy ordering},\ }\href
  {https://doi.org/https://doi.org/10.1021/ic501002j} {\bibfield  {journal}
  {\bibinfo  {journal} {Inorg. Chem.}\ }\textbf {\bibinfo {volume} {54}},\
  \bibinfo {pages} {722–732} (\bibinfo {year} {2015})}\BibitemShut {NoStop}%
\bibitem [{\citenamefont {Bao}\ \emph {et~al.}(2021)\citenamefont {Bao},
  \citenamefont {Zheng}, \citenamefont {Wen}, \citenamefont {Ramakrishnan},
  \citenamefont {Zheng}, \citenamefont {Jiang}, \citenamefont {Bugaris},
  \citenamefont {Cao}, \citenamefont {Chung}, \citenamefont {van Smaalen},\
  and\ \citenamefont {Kanatzidis}}]{Bao2021}%
  \BibitemOpen
  \bibfield  {author} {\bibinfo {author} {\bibfnamefont {J.-K.}\ \bibnamefont
  {Bao}}, \bibinfo {author} {\bibfnamefont {H.}~\bibnamefont {Zheng}}, \bibinfo
  {author} {\bibfnamefont {J.}~\bibnamefont {Wen}}, \bibinfo {author}
  {\bibfnamefont {S.}~\bibnamefont {Ramakrishnan}}, \bibinfo {author}
  {\bibfnamefont {H.}~\bibnamefont {Zheng}}, \bibinfo {author} {\bibfnamefont
  {J.~S.}\ \bibnamefont {Jiang}}, \bibinfo {author} {\bibfnamefont
  {D.}~\bibnamefont {Bugaris}}, \bibinfo {author} {\bibfnamefont
  {G.}~\bibnamefont {Cao}}, \bibinfo {author} {\bibfnamefont {D.~Y.}\
  \bibnamefont {Chung}}, \bibinfo {author} {\bibfnamefont {S.}~\bibnamefont
  {van Smaalen}},\ and\ \bibinfo {author} {\bibfnamefont {M.~G.}\ \bibnamefont
  {Kanatzidis}},\ }\bibfield  {title} {\bibinfo {title} {Superconductivity in
  {Y}$_4${Ru}{Ge}$_8$ with a vacancy-ordered {Ce}{Ni}{Si}$_2$‑type
  superstructure},\ }\href
  {https://doi.org/https://doi.org/10.1021/acs.chemmater.1c02488} {\bibfield
  {journal} {\bibinfo  {journal} {Chem. Mater.}\ }\textbf {\bibinfo {volume}
  {33}},\ \bibinfo {pages} {7839–7847} (\bibinfo {year} {2021})}\BibitemShut
  {NoStop}%
\bibitem [{\citenamefont {Akhiezer}\ \emph {et~al.}(1961)\citenamefont
  {Akhiezer}, \citenamefont {Baryakhtar},\ and\ \citenamefont
  {Kazanov}}]{Akhiezer1961}%
  \BibitemOpen
  \bibfield  {author} {\bibinfo {author} {\bibfnamefont {A.~I.}\ \bibnamefont
  {Akhiezer}}, \bibinfo {author} {\bibfnamefont {V.~G.}\ \bibnamefont
  {Baryakhtar}},\ and\ \bibinfo {author} {\bibfnamefont {M.~I.}\ \bibnamefont
  {Kazanov}},\ }\bibfield  {title} {\bibinfo {title} {Spin waves in
  ferromagnets and antiferromagnets},\ }\href
  {https://doi.org/10.1070/PU1961v003n04ABEH003309} {\bibfield  {journal}
  {\bibinfo  {journal} {Sov. Phys. Uspekhi}\ }\textbf {\bibinfo {volume} {3}},\
  \bibinfo {pages} {567} (\bibinfo {year} {1961})}\BibitemShut {NoStop}%
\bibitem [{\citenamefont {Guillaume}(1897)}]{Guillaume1897}%
  \BibitemOpen
  \bibfield  {author} {\bibinfo {author} {\bibfnamefont {C.~E.}\ \bibnamefont
  {Guillaume}},\ }\bibfield  {title} {\bibinfo {title} {Recherches sur les
  aciers au nickel. {D}ilatations aux temperatures elevees; resistance
  electrique},\ }\href@noop {} {\bibfield  {journal} {\bibinfo  {journal} {CR
  Acad. Sci.}\ }\textbf {\bibinfo {volume} {125}},\ \bibinfo {pages} {235}
  (\bibinfo {year} {1897})}\BibitemShut {NoStop}%
\bibitem [{\citenamefont {Klimczuk}\ \emph {et~al.}(2012)\citenamefont
  {Klimczuk}, \citenamefont {Walker}, \citenamefont {Springell}, \citenamefont
  {Shick}, \citenamefont {Hill}, \citenamefont {Gaczy{\'n}ski}, \citenamefont
  {Gofryk}, \citenamefont {Kimber}, \citenamefont {Ritter}, \citenamefont
  {Colineau}, \citenamefont {Griveau}, \citenamefont {Bou{\"e}xi{\`e}re},
  \citenamefont {Eloirdi}, \citenamefont {Cava},\ and\ \citenamefont
  {Caciuffo}}]{Klimczuk2012}%
  \BibitemOpen
  \bibfield  {author} {\bibinfo {author} {\bibfnamefont {T.}~\bibnamefont
  {Klimczuk}}, \bibinfo {author} {\bibfnamefont {H.~C.}\ \bibnamefont
  {Walker}}, \bibinfo {author} {\bibfnamefont {R.}~\bibnamefont {Springell}},
  \bibinfo {author} {\bibfnamefont {A.~B.}\ \bibnamefont {Shick}}, \bibinfo
  {author} {\bibfnamefont {A.~H.}\ \bibnamefont {Hill}}, \bibinfo {author}
  {\bibfnamefont {P.}~\bibnamefont {Gaczy{\'n}ski}}, \bibinfo {author}
  {\bibfnamefont {K.}~\bibnamefont {Gofryk}}, \bibinfo {author} {\bibfnamefont
  {S.~A.~J.}\ \bibnamefont {Kimber}}, \bibinfo {author} {\bibfnamefont
  {C.}~\bibnamefont {Ritter}}, \bibinfo {author} {\bibfnamefont
  {E.}~\bibnamefont {Colineau}}, \bibinfo {author} {\bibfnamefont {J.-C.}\
  \bibnamefont {Griveau}}, \bibinfo {author} {\bibfnamefont {D.}~\bibnamefont
  {Bou{\"e}xi{\`e}re}}, \bibinfo {author} {\bibfnamefont {R.}~\bibnamefont
  {Eloirdi}}, \bibinfo {author} {\bibfnamefont {R.~J.}\ \bibnamefont {Cava}},\
  and\ \bibinfo {author} {\bibfnamefont {R.}~\bibnamefont {Caciuffo}},\
  }\bibfield  {title} {\bibinfo {title} {Negative thermal expansion and
  antiferromagnetism in the actinide oxypnictide {Np}{Fe}{As}{O}},\ }\href
  {https://doi.org/10.1103/PhysRevB.85.174506} {\bibfield  {journal} {\bibinfo
  {journal} {Phys. Rev. B}\ }\textbf {\bibinfo {volume} {85}},\ \bibinfo
  {pages} {174506} (\bibinfo {year} {2012})}\BibitemShut {NoStop}%
\bibitem [{\citenamefont {Jaime}\ \emph {et~al.}(2017)\citenamefont {Jaime},
  \citenamefont {Saul}, \citenamefont {Salamon}, \citenamefont {Zapf},
  \citenamefont {Harrison}, \citenamefont {Durakiewicz}, \citenamefont
  {Lashley}, \citenamefont {Andersson}, \citenamefont {Stanek}, \citenamefont
  {Smith},\ and\ \citenamefont {Gofryk}}]{Jaime2017}%
  \BibitemOpen
  \bibfield  {author} {\bibinfo {author} {\bibfnamefont {M.}~\bibnamefont
  {Jaime}}, \bibinfo {author} {\bibfnamefont {A.}~\bibnamefont {Saul}},
  \bibinfo {author} {\bibfnamefont {M.}~\bibnamefont {Salamon}}, \bibinfo
  {author} {\bibfnamefont {V.~S.}\ \bibnamefont {Zapf}}, \bibinfo {author}
  {\bibfnamefont {N.}~\bibnamefont {Harrison}}, \bibinfo {author}
  {\bibfnamefont {T.}~\bibnamefont {Durakiewicz}}, \bibinfo {author}
  {\bibfnamefont {J.~C.}\ \bibnamefont {Lashley}}, \bibinfo {author}
  {\bibfnamefont {D.~A.}\ \bibnamefont {Andersson}}, \bibinfo {author}
  {\bibfnamefont {C.~R.}\ \bibnamefont {Stanek}}, \bibinfo {author}
  {\bibfnamefont {J.~L.}\ \bibnamefont {Smith}},\ and\ \bibinfo {author}
  {\bibfnamefont {K.}~\bibnamefont {Gofryk}},\ }\bibfield  {title} {\bibinfo
  {title} {Piezomagnetism and magnetoelastic memory in uranium dioxide},\
  }\href {https://doi.org/10.1038/s41467-017-00096-4} {\bibfield  {journal}
  {\bibinfo  {journal} {Nat. Commun.}\ }\textbf {\bibinfo {volume} {8}},\
  \bibinfo {pages} {99} (\bibinfo {year} {2017})}\BibitemShut {NoStop}%
\bibitem [{\citenamefont {Shrestha}\ \emph {et~al.}(2017)\citenamefont
  {Shrestha}, \citenamefont {Antonio}, \citenamefont {Jaime}, \citenamefont
  {Harrison}, \citenamefont {Mast}, \citenamefont {Safarik}, \citenamefont
  {Durakiewicz}, \citenamefont {Griveau},\ and\ \citenamefont
  {Gofryk}}]{Shrestha2017}%
  \BibitemOpen
  \bibfield  {author} {\bibinfo {author} {\bibfnamefont {K.}~\bibnamefont
  {Shrestha}}, \bibinfo {author} {\bibfnamefont {D.}~\bibnamefont {Antonio}},
  \bibinfo {author} {\bibfnamefont {M.}~\bibnamefont {Jaime}}, \bibinfo
  {author} {\bibfnamefont {N.}~\bibnamefont {Harrison}}, \bibinfo {author}
  {\bibfnamefont {D.~S.}\ \bibnamefont {Mast}}, \bibinfo {author}
  {\bibfnamefont {D.}~\bibnamefont {Safarik}}, \bibinfo {author} {\bibfnamefont
  {T.}~\bibnamefont {Durakiewicz}}, \bibinfo {author} {\bibfnamefont {J.-C.}\
  \bibnamefont {Griveau}},\ and\ \bibinfo {author} {\bibfnamefont
  {K.}~\bibnamefont {Gofryk}},\ }\bibfield  {title} {\bibinfo {title}
  {Tricritical point from high-field magnetoelastic and metamagnetic effects in
  {UN}},\ }\href {https://doi.org/10.1038/s41598-017-06154-7} {\bibfield
  {journal} {\bibinfo  {journal} {Sci. Rep.}\ }\textbf {\bibinfo {volume}
  {7}},\ \bibinfo {pages} {6642} (\bibinfo {year} {2017})}\BibitemShut
  {NoStop}%
\bibitem [{\citenamefont {Blundell}(2001)}]{Blundell2001}%
  \BibitemOpen
  \bibfield  {author} {\bibinfo {author} {\bibfnamefont {S.}~\bibnamefont
  {Blundell}},\ }\href@noop {} {\emph {\bibinfo {title} {Magnetism in Condensed
  Matter}}}\ (\bibinfo  {publisher} {Oxford University Press},\ \bibinfo
  {address} {New York, NY, USA},\ \bibinfo {year} {2001})\ \bibinfo {note}
  {(part of: Oxford Master Series in Condensed Matter Physics)}\BibitemShut
  {NoStop}%
\bibitem [{\citenamefont {Tro\'{c}}\ \emph {et~al.}(2016)\citenamefont
  {Tro\'{c}}, \citenamefont {Samsel-Czeka{\l}a}, \citenamefont {Pikul},
  \citenamefont {Andreev}, \citenamefont {Gorbunov}, \citenamefont {Skourski},\
  and\ \citenamefont {Sznajd}}]{Troc2016}%
  \BibitemOpen
  \bibfield  {author} {\bibinfo {author} {\bibfnamefont {R.}~\bibnamefont
  {Tro\'{c}}}, \bibinfo {author} {\bibfnamefont {M.}~\bibnamefont
  {Samsel-Czeka{\l}a}}, \bibinfo {author} {\bibfnamefont {A.}~\bibnamefont
  {Pikul}}, \bibinfo {author} {\bibfnamefont {A.~V.}\ \bibnamefont {Andreev}},
  \bibinfo {author} {\bibfnamefont {D.~I.}\ \bibnamefont {Gorbunov}}, \bibinfo
  {author} {\bibfnamefont {Y.}~\bibnamefont {Skourski}},\ and\ \bibinfo
  {author} {\bibfnamefont {J.}~\bibnamefont {Sznajd}},\ }\bibfield  {title}
  {\bibinfo {title} {Electronic structure of {UN} based on specific heat and
  field-induced transitions up to 65 {T}},\ }\href
  {https://doi.org/10.1103/PhysRevB.94.224415} {\bibfield  {journal} {\bibinfo
  {journal} {Phys. Rev. B}\ }\textbf {\bibinfo {volume} {94}},\ \bibinfo
  {pages} {224415} (\bibinfo {year} {2016})}\BibitemShut {NoStop}%
\bibitem [{\citenamefont {Perricone}(2002)}]{Perricone2002}%
  \BibitemOpen
  \bibfield  {author} {\bibinfo {author} {\bibfnamefont {A.}~\bibnamefont
  {Perricone}},\ }\href@noop {} {} (\bibinfo {year} {2002}),\ \bibinfo {note}
  {{S}ynth{\`e}se, structures cristallines et propri{\'e}t{\'e}s
  magn{\'e}tiques de nouveaux compos{\'e}s interm{\'e}talliques dans le
  syst{\`e}me binaire U-Ni et les syst{\`e}mes ternaires U-Ni-X (X: Ge, Si,
  Al), Doctoral dissertation (in {F}rench), {U}niversit{\'e} {R}ennes 1,
  Rennes, France}\BibitemShut {NoStop}%
\end{thebibliography}
\end{document}